\pdfoutput=1
\documentclass{article}

\usepackage[english]{babel}
\usepackage[utf8]{inputenc}
\usepackage{color}

\usepackage{graphicx} 
\usepackage{amsmath} 
\usepackage{amssymb} 
\usepackage{amsthm} 
\usepackage{tabularx} 
\usepackage{siunitx} 
\usepackage{algorithm} 
\usepackage[noend]{algpseudocode} 
\usepackage{caption} 
\usepackage{subcaption} 
\usepackage{filecontents} 

\usepackage{authblk}

\everymath{\displaystyle}
\graphicspath{{img/}}

\title{A novel highly efficient Lagrangian model for massively multidomain simulations: parallel context}

\author[1]{Sebastian~Florez\thanks{corresponding author}}
\author[1]{Julien~Fausty}
\author[1]{Karen~Alvarado}
\author[1]{Brayan~Murgas}
\author[1]{Marc~Bernacki}

\affil[1]{Mines-ParisTech, PSL-Research University, CEMEF – Centre de mise en forme des mat\'{e}riaux, CNRS UMR 7635, CS 10207 rue Claude Daunesse, 06904 Sophia Antipolis Cedex, France}%

\begin{document}
\maketitle

\section*{abstract}

A new method for the simulation of evolving multi-domains problems has been introduced in a previous work (RealIMotion), Florez et al. (2020). In this article further developments of the model will be presented. The main focus here is a robust parallel implementation using a distributed-memory approach with the Message Passing Interface (MPI) library OpenMPI. The original 2D sequential methodology consists in a modified front-tracking approach where the main originality is that not only interfaces between domains are discretized but their interiors are also meshed. The interfaces are tracked based on the topological degree of each node on the mesh and the remeshing and topological changes of the domains are driven by selective local operations performed over an element patch. The accuracy and the performance of the sequential method has proven very promising in Florez et al. (2020). In this article a parallel implementation will be discussed and tested in context of motion by curvature flow for polycrystals, i.e. by considering Grain Growth (GG) mechanism. Results of the performance of the model are given and comparisons with other approaches in the literature are discussed.


\section{Introduction}

The simulation of the dynamic of massive multidomain problems have been addressed by several numerical approaches \cite{Anderson1984, Srolovitz1984, Hesselbarth1991, Kawasaki1989, Brakke1992}. Usually, these approaches aim to enhance the accuracy and the performance of the developed framework by means of computational optimizations, especially when the model intends to simulate thousands of domains. In the context of modeling the microstructure of metallic materials, many numerical frameworks exist that are capable of handling thousands of domains: the probabilistic \textbf{Monte Carlo} (MC) method (or Pott's model) \cite{Anderson1984, Srolovitz1984, Srolovitz1988, Rollett1992, Peczak1993a, Peczak1995, Radhakrishnan1995} and the \textbf{Cellular Automata} (CA) method \cite{Hesselbarth1991, Pezzee1994, Liu1996, Raabe1998, Raabe1999, Rollett2001, Lin2016, Hallberg2010, Rauch2015, Li2016, Madej2018} both relying in a computational domain discretized in voxels, making them very efficient in a parallel context \cite{Rauch2015, Madej2018} although a greater effort must be done for applications where the deformation of the domain is expected \cite{Madej2018}. Furthermore, the \textbf{Level-Set} (LS) method \cite{Osher1988, Merriman1994} coupled with a \textbf{Finite Element} (FE) framework (FE-LS) \cite{loge2008, Bernacki2008,Cruz-Fabiano2014,Scholtes2016b, Florez2020} or with a \textbf{Fast-Fourier Transformation} (FFT) framework (FFT-LS) \cite{Miesen2017, Elsey2009, Elsey2013} are also examples of robust and highly efficient methods in this context, where the FFT-LS method is also limited by the use of regular grids but accounts for a better computational performance. These both rely on the definition of each domain by the use of signed LS functions to perform their computations allowing the definition of \emph{sharp interfaces}\footnote{Interfaces are defined as the zero isovalue of the LS function.} of domains, while their evolution is in general formulated via a transport equation. Another model with a high affinity with the LS model is the \textbf{Phase-Field} (PF) method \cite{ChenPF1994, Fan1997, Krill2002, Kazaryan2001, Moelans2008, Moelans2013, Chen2015, Chang2017, Steinbach1996, Garcke1999, Miyoshi2017, Takaki2014} where the main difference with the LS method is the use of \emph{diffuse interfaces}\footnote{Interfaces are contained within a volume in 3D or surface in 2D, defined with a numeric field (order parameter of phase-field) which varies smoothly in the range [0,1] or [-1,1].}, their evolution is formulated based on the minimisation of the free energy of the system. Aditionally to the MC, CA, LS and PF methods, models based on the Lagrangian displacement of interfaces can be used as in \textbf{vertex} approaches \cite{Soares1985, Kawasaki1989, Weygand1998, Lepinoux2010, BarralesMora2010, Mellbin2015} or \textbf{Front-Tracking} approaches \cite{Frost1988, Brakke1992, Couturier2003, Couturier2004, Couturier2005, Becker2008} which have been taken as inspiration for the developement of a new front-tracking method topological REmeshing in lAgrangian framework for Large Interface MOTION (RealIMotion hereafter TRM) introduced in a previous publication \cite{Florez2020b}.\\

The initial algorithm proposed in \cite{Florez2020b} described a series of selective remeshing operations strongly influenced by the works in \cite{Compere2008, Compere2009} and a data structure adapted specifically to multidomain problems. All remeshing operations where performed over a local patch of elements while the data structure was based on the local topology (of the simulated domain) that each node represents (multiple junction or \emph{Point}, grain boundary or \emph{Line}, grain bulk or \emph{Surface}). Geometric properties of the interface were computed with the help of piece-wise polynomials (Natural Splines) and the movement of interfaces was based on a Lagrangian model. Moreover, some topological changes on the structure of the multidomain problem were addressed by means of local remeshing operations, where the Node-Collapse algorithm was used to treat the disappearance of domains while the Point-Splitting algorithm allowed the creation of new interfaces.\\

The performance of the TRM model was tested and compared against a classical front capturing LS-FE framework \cite{Bernacki2008,Cruz-Fabiano2014,maire2016} in an isotropic GG context. Multiple test cases were performed concluding in an improvement of the accuracy and performance when using the TRM model. A 14 times reduction in the computational cost of the TRM model was observed as compared to the best case scenario of the LS-FE framework \cite{Florez2020}.\\

Even though the results obtained in \cite{Florez2020b} were promising, the TRM model was lacking a very important component in order to make a more accurate comparison in terms of computational performance : a parallel implementation. Two categories of parallel frameworks can be designed in order to use all the capabilities offered by modern computational units. These two categories are differentiated by the management of the active memory of the running processes: \emph{shared memory}, where all processors share and interact with the same location in memory, and \emph{distributed memory}, where each process has its own independent memory location. The choice of whether to use one or the other relies strongly on the hardware architecture on which the model is intented to run. Normally an application with a shared memory framework can not be used over a supercomputer cluster with thousands of cores, as the memory in these systems is not connected to a single board but it is distributed between several independent CPUs connected to the same network. In other words, shared memory can only be used in single machines while distributed memory is intended to be used both whithin single machines and over a network of interconnected CPUs. Each processor in a distributed memory system needs a way to communicate information to other processors. This communication can be established by the use of standard protocols such as MapReduce \cite{Dean2004} or the Message Passing Interface (MPI) \cite{Walker1992}. These methods of sending informations between processors cause some overhead on the global multi-processor application, hence obtaining a lower performance than when using a shared memory approach. On the other hand, shared memory protocols can also add some overhead when implementing an environment safe of race conditions: when two or more processors try to write to the same memory location at the same time. Here, a distributed memory approach using the MPI protocol in order to address the parallel implementation of the TRM model is proposed to enable a broader range of hardware compatibility.\\

Very few publications exist in the litterature regarding the parallel implementation of Front-Tracking models, examples of these works can be found in \cite{Laucoin1996, DaSilveiraNeto2016} in the context of two-phase flows, and in \cite{Pan2012} in a more general context of multiphase flows, however tested for only just a few domains (3 domains). Similarly, examples of vertex models using a parallel scheme can be found in \cite{Sussman2017, Madhikar2018} using a GPU-based parallel approach (hence using a \emph{shared memory} approach) in the context of molecular-dynamics simulations. These examples, although very impressive, are considerably different from the TRM approach, as an important strength of the proposed methodology is to deal with unstructured finite element meshes, allowing (i) the discretization of domain boundaries by more than a vertex-vertex connection \cite{Sussman2017, Madhikar2018}, (ii) the direct evolution/migration of the nodes of the mesh as a mean of boundary kinetics, without the use of two discretization approaches \cite{Laucoin1996, DaSilveiraNeto2016} and (iii) large deformation modeling (thus adapted to the context of recrystallization modeling in hot metal forming)  which is generally not accessible to regular grid approaches (MC, CA, and FFT models).\\

In this paper the parallel implementation of the TRLM model will be presented and tested in multiple hardware settings. The special algorithms to address the parallel framework will be explained. Moreover, an additional tool is needed when performing parallel computations over a distributed memory approach, in order to solve mesh-based problems: the initial partitioning of the numerical domain and the redistribution (when necessary) of charges (repartitioning) through the evolution of the simulation. Here we opted to use the open source library Metis \cite{Karypis1998} in order to obtain the initial partitioning while the redistribution algorithm has been developed and will be also presented in this paper.\\

Performance and speed-up of the model will be given and compared to other highly efficient parallel methods in the litterature in the context of GG \cite{Miesen2017, Elsey2013}. Also, in the following, the term {\it domain} will reference an individual grain in a microstructure, however, the parallel TRM approach can be extended to any massively multi-domain problem immersed in a 2D triangular mesh.\\

\section{The TRM model : sequential approach in a GG context} \label{sec:theTRMmodel}

In \cite{Florez2020b}, a numerical method for the TRM model was presented, this numerical method is built on a data structure defining the current state of the multidomain framework, defining geometrical entities such as \emph{points}, \emph{lines} and \emph{surfaces}. Each point is composed of a P-Node (defining a node of the mesh with a topological degree equal to 0) and a set of connections to other points and lines. Each line is defined by an ordered set of L-Nodes (nodes with a topological degree equal to 1), an initial point and a final point. Finally, surfaces are defined by a set of S-Nodes (nodes with a topology degree equal to 2), a set of elements and a set of delimiting lines and points.\\

\begin{figure}[!h]
\centering
\includegraphics[width=1.0\textwidth] {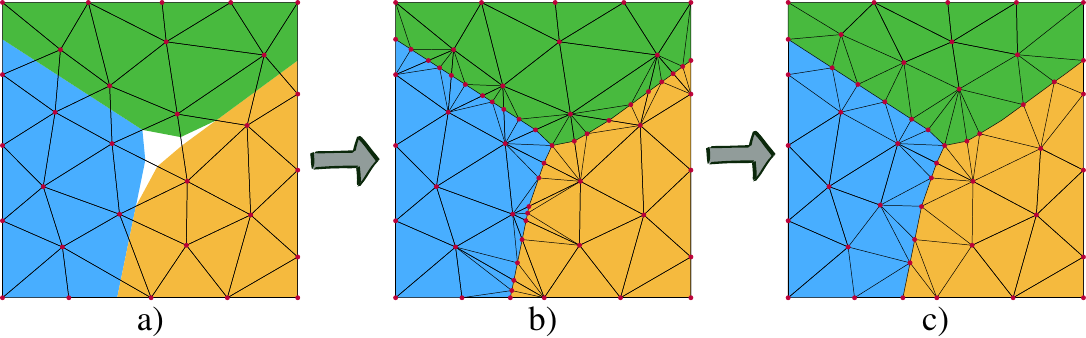}
\caption{Example of the behaviour of the \emph{fitting and joining} procedure presented in \cite{Florez2020} when performed over a triangular mesh with 3 LS fields interpolated linearly, each color represent a phase. a) initial state, interfaces are implicitly described, b) state after applying the \emph{fitting and joining} algorithm, nodes appear at the interface between 2 phases and at the multiple junction. Vacuum regions dissapear and all elements belong to one phase. c) Resulting mesh after the mesh adaptation procedure \cite{Florez2020}. }
\label{fig:FitJoinAlgo}
\end{figure}

Once the data structure was defined, a preprocessor of the TRM model was introduced. This preprocessor makes an interface between the LS domain definition and the TRM data structure based on the works presented in \cite{Florez2020, Shakoor2017ijnme}, where an initially implicit mesh (with an immersed LS data set as in \cite{Scholtes2015, Scholtes2016}) is transformed into a body-fitted mesh via a \emph{joining and fitting algorithm}, where all limits of domains are explicitly defined by some of the nodes of the mesh (see Fig. \ref{fig:FitJoinAlgo} for an illustration). Subsequently, four algorithms for the reconstruction of the domains for the the TRM model were presented: Nodal geometric tagging, Point Reconstruction, Line Reconstruction and Surface Reconstruction. These algorithms completly define the data structure of the TRM model of a LS data set immersed into a body-fitted mesh. It is important to emph that from this point forward, the LS data set is no longer of use for our model, and that all geometric properties of the interfaces and domains are computed using purely geometric approximations built upon the data structure of the TRM model. Natural parametric splines \cite{DeBoor1978} are used to approximate the domain interfaces (lines) with third degree piece-wise polynomials, local geometric properties of the interfaces such as the curvature $\kappa$ and the normal $\vec{n}$, are deducted from these approximations.\\

Once a velocity has been computed on the mesh, dependending on the physical model being simulated, each node $N_i$ of the mesh is moved to a new position $\vec{r_i}$ in a Lagrangian way using the velocity vector field $\vec{v}$ and a given time step $dt$.

\begin{equation}
\label{Eq:lagrangianDisplacement}
\centering
\vec{r_i}=\vec{r_{i}^0} + \vec{v_i}~dt. 
\end{equation}

where $\vec{r_{i}^0}$ is the position of the node $N_i$ before its displacement. This step is algorithmically independent from the physics of the problem. Mesh conformity (in a FE sense) is ensured by a local-iteratively movement-halving algorithm, reducing at each iteration the movement by half when an invalid mesh configuration was encountered i.e. an element flipping (see \cite{Florez2020b}).\\

TRM model also involves a particular remeshing procedure. The TRM data structure needs to be maintained at all times during remeshing to ensure all geometric computations over the defined geometric entities. When a remeshing procedure is performed, the mesh evolves and the sets defining each geometric entity have to adapt, therefore, the remeshing procedure must take into account the local data structure of the nodes and elements involved in each remeshing operation. The remeshing strategy of the TRM model uses the separate definition of locally \emph{selective}\footnote{The word selective denotes a variation of the original remeshing operations when performed over the data structure of the TRM model, as each remeshing operation will be performed differently over nodes with different topology (P-Node, L-Node and S-Node)} remeshing operators: selective vertex smoothing, selective node collapsing, selective edge splitting, selective edge swapping and selective vertex gliding, see \cite{Florez2020b} for a complete definition of each operator. As a general rule, the remeshing procedure is performed to increase the general quality $Q$ of the mesh (or a patch of elements of the mesh). Here, the notion of mesh quality $Q$ will be computed as a factor of the shape and the size of the elements using the same approach as in \cite{Shakoor2015}. However, the selective remeshing procedure is not only driven by the local mesh quality $Q$, but also by the local topological degree of the nodes involved in the operation. In \cite{Florez2020b} a global remeshing procedure was introduced, driven by two nodal fields $\delta_c$ and $\delta_s$ corresponding to the collapsing and splitting fields, and a minimum quality shape coefficient $q_s$. The complete remeshing procedure is summarized in Algorithm \ref{alg:Remeshing}.

\begin{algorithm}
\caption{Remeshing algorithm \cite{Florez2020b}}\label{alg:Remeshing}
\begin{algorithmic}[1]
\ForAll{Nodes : $N_i$}
	\ForAll{Neighbors of $N_i$ : $N_j$}
		\If{$\delta_c(N_i,N_j)<|\overline{N_iN_j}|$} 
			\State selective node collapse : $N_j$ collapses $N_i$
		\EndIf
	\EndFor
\EndFor
\ForAll{S-Nodes : $SN_i$}
    \State selective vertex smoothing : $SN_i$
\EndFor
\ForAll{L-Nodes : $LN_i$}
	\State selective vertex gliding : $LN_i$
\EndFor
\ForAll{Edges : $\lbrace{N_i, N_j\rbrace}_{j>i}$}
	\If{$\delta_s(N_i,N_j)>|\overline{N_iN_j}|$} 
		\State selective edge splitting : $N_i, N_j$
	\EndIf
\EndFor
\ForAll{Elements with $Q_s<q_s$ : $E_i$}
	\ForAll{Edges of $E_i$: $\lbrace N_j, N_k\rbrace_{k>j} $ }
		\If{quality $Q_{mean}(N_j, N_k)$ will improve  by swapping}
		\State selective edge swapping : $\lbrace N_j, N_k\rbrace$
		\EndIf
	\EndFor
\EndFor
\end{algorithmic}
\end{algorithm}

\subsection{Grain Growth Modeling}

The simulation of microstructural evolutions are given by the addition of complex and different mechanisms such as GG \cite{Bernacki2008,Cruz-Fabiano2014,maire2016,furstoss2018}, Recrystallization (ReX)  \cite{Bernacki2009, Bernacki2011, Scholtes2016, Maire2017}  or Zener Pinning (ZP) \cite{Weygand1999, Couturier2003, Couturier2004, Couturier2005}. In \cite{Florez2020b} isotropic GG was used to compare the TRM model to other approaches (LS-FE \cite{Bernacki2011,Cruz-Fabiano2014,maire2016}). The base model used to represent this mechanism is commonly known as curvature flow, where the velocity of an interface is proportional to its local mean curvature.

During grain growth the velocity $\vec{v}$ at every point of the interface can be approximated by the following equation:
\begin{equation}
\label{Eq:VelocityEquation}
\centering
\vec{v}=-M \gamma\kappa \vec{n},
\end{equation}
where $M$ is the mobility of the interface, $\gamma$ the grain boundary energy,  $\kappa$ the local curvature in 2D and trace of the curvature tensor in 3D and $\vec{n}$ the outward unit normal to the grain interface. Isotropic conditions will be considered hereafter: $M$ is assumed only dependent on the temperature, which is constant over the space and $\gamma$ is constant.

Note that the velocity at multiple junctions can not be deducted from Eq. \ref{Eq:VelocityEquation} as the curvature and normal at these points can not be mathematically computed. An alternative approach is needed: Model II of \cite{Kawasaki1989}, where the product $\kappa \vec{n}$ is deducted from an approximation of the free energy equation of the whole system in a vertex context is used.

The presented algorithm can be modified in order to simulate other different multi-domain problems by adapting the velocity equation defined in Eq \ref{Eq:VelocityEquation}. Multiple topological changes of the polycrystal structure (grain disappearance and quadruple point dissociation) usually encountered in GG will also be taken into account by the modeling algorithm.\\

\subsection{Topological changes during grain growth}

The first topological change observed during GG is the collapse of an entity with a topological degree different from 0 (a line, a surface or a volume) to a point. Grain disappearance, for example, is the result of the multiple topological collapses on the polycrystal structure where all interfaces of a grain successively collapse into multiple junctions (points where more than two interfaces intersect), eventually reducing the volume of the grain to a single point. The TRM model handles this topological event during the remeshing process where the node collapsing strategy successively collapse nodes that are too close. Note that P-Nodes have a certain predominance over all other types of nodes when using the selective node collapsing algorithm of section 2.4.1 of \cite{Florez2020b}. This means that when a phase disappears, a point of its interface will always remain at the location of the event.

\begin{figure}[!h]
\centering
\includegraphics[width=1.0\textwidth] {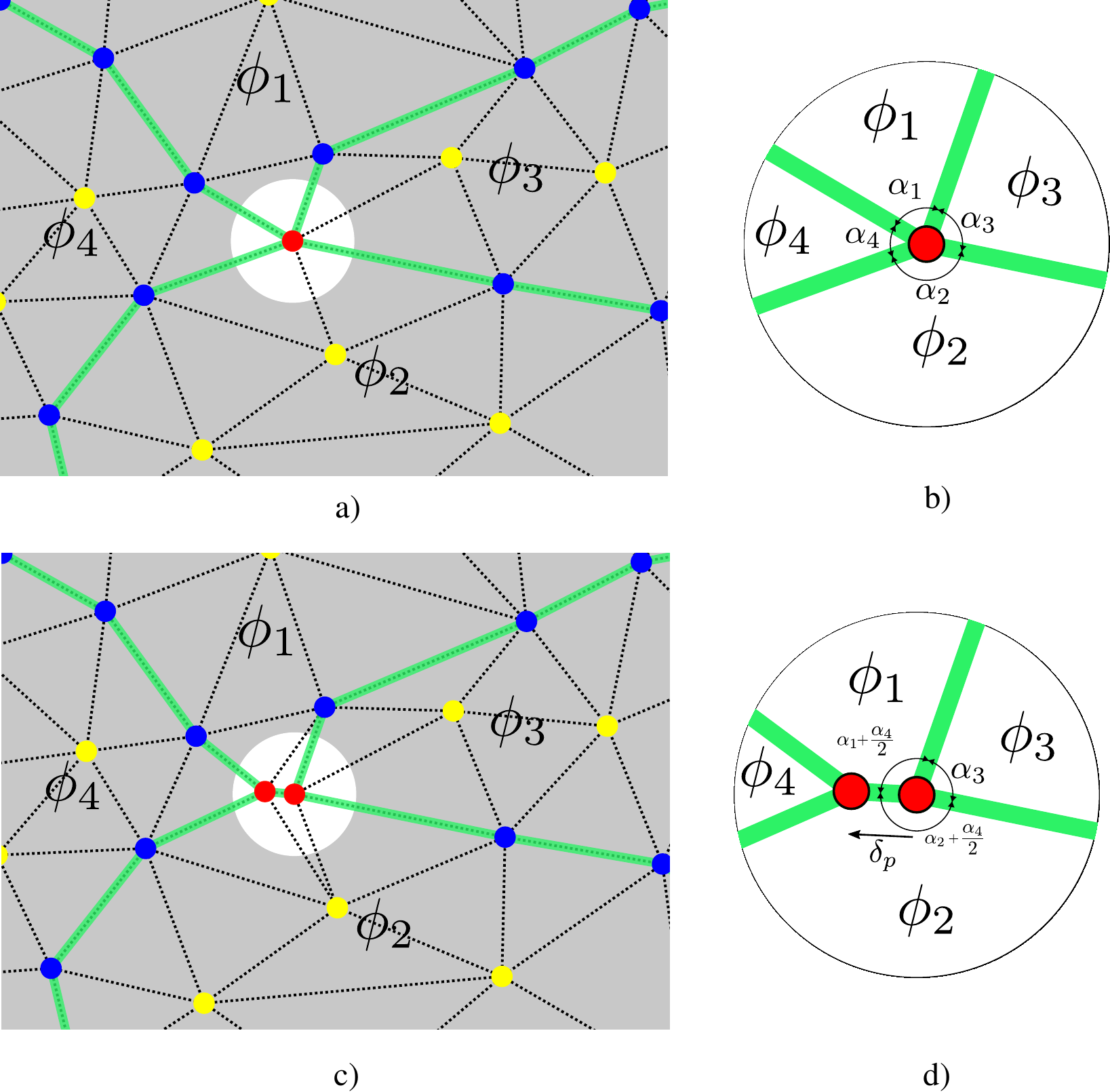}
\caption{Example of a multiple junction decomposition with isotropic boundary energies, a) initial state. b) detailed view of the multiple junction. Here $\alpha_4<\alpha_1<\alpha_3<\alpha_2$, the choice of the phase to detach from the multiple point is determinated by the lowest angle. c) final state after the decomposition procedure, two new elements are created.  d) detailed view of the decomposition, interfaces of $\phi_4$ are separated from the initial multiple junction and a new interface is created ($\phi_1$ - $\phi_2$), the new point will be placed along the line determined by the angle $\alpha_4/2$ measured from one of the two detached interface to the inner side of $\phi_4$ at a distance $\delta_p$ from the initial multiple point \cite{Florez2020b}.}
\label{fig:SplitCuadruplePointIso}
\end{figure}

The remeshing procedure is capable of handling some of the topological changes on the structural domain: Line dissappearence and Surface dissappearence, which are both a product of the selective node collapse. However, one special topological change can not be reproduced by the remeshing algorithm alone: the decomposition of an unstable multiple point into stable multiple points (e.g. triple points in the context of isotropic GG) and lines; this topological operation needs an special operator that is not classified as a remeshing procedure, because it is derived purely by the instability produced by the current configuration. In \cite{Florez2020b}, it was explained that the decomposition process in an anisotropic context where the surface energy depends on variables such as the crystallographic orientation and interface inclination is very complex as there may exist multiple solutions that reduce the total surface energy, but only the configuration with the lowest energy should be considered. However, in an isotropic context, the solution to the decomposition problem is only dependent on the angles formed between each line connected to the initial unstable multiple point; the solution is then to separate the pair of lines forming the lower angle in the initial configuration, creating a line from the initial multiple point to the separated vertex between the pair of lines as explained in figure \ref{fig:SplitCuadruplePointIso}(b and c)), figure \ref{fig:SplitCuadruplePointIso}(a and b)) which illustrate the changes of the mesh during the decomposition implemented on the TRM model.

\subsection{the TRM algorithm for Grain Growth}

The complete algorithm for an increment of the TRM model is presented in Algorithm \ref{alg:TRMGG}.

\begin{algorithm}
\caption{Isotropic Grain Growth TRM Algorithm}\label{alg:TRMGG}
\begin{algorithmic}[1]
\State Perform \textbf{Remeshing Algorithm} (Algorithm \ref{alg:Remeshing})
\ForAll{Points: $P_i$}
	\While{Number of Connections $>3$ }
		\State split multiple point $P_i$.
	\EndWhile
\EndFor
\ForAll{Lines : $L_i$}
    \State Compute the natural spline approximation of $L_i$.
\EndFor
\ForAll{L-Nodes : $LN_i$}
    \State Compute curvature and normal ($\kappa \vec{n}$) over $LN_i$.
\EndFor
\ForAll{P-Nodes : $PN_i$}
    \State Compute the product $\kappa \vec{n}$ over $PN_i$ using model II of \cite{Kawasaki1989}.
\EndFor
\ForAll{L-Nodes and P-Nodes : $LPN_i$}
	\State Compute velocity $\vec{v_i}$ of Node $LPN_i$
    \State Iterative movement with flipping check over $LPN_i$
\EndFor
\end{algorithmic}
\end{algorithm}

\section{Parallel strategy for the TRM model}

The TRM model introduced in \cite{Florez2020b} lacks an important component to be a competitive tool to the more classical methods presented in the literature to model microstructural evolutions, this component is the ability to perform large computations thanks to a parallel implementation.\\

As mentioned in the introduction, in order to address the parallel implementation of the TRM model a \emph{distributed memory} approach  is proposed, using the standard communication protocol ``Message Passing Interface" (MPI) \cite{Walker1992} to communicate information between processes.\\

Regardless of the choice of the memory management of the parallel framework (\emph{shared memory} or \emph{distributed memory} system), two additional tools are needed when performing parallel computations in order to solve mesh-based problems: the initial partitioning of the numerical domain and the redistribution of charges (repartitioning) through the evolution of the simulation. These two tools are essential to ensure an equilibrium to the memory and to the charge that each processor has to handle. In a \emph{distributed memory} system, the whole numerical domain can be divided to give to each processor one part of that domain to handle at the begining of the simulation. Multiple tools already exist in order to partition a mesh made of simplices, normally these tools can work with topologies much more complex than an Eulerian mesh, as they are engineered to treat graphs\footnote{mathematical structures used to model pairwise relations (edges) between objects (vertices, nodes or points)}. Here we opted to use the free library Metis \cite{Karypis1998} in order to obtain the initial partitioning.\\

Once the partition of the domain is performed and a strategy for the re-equilibrium of charges is developed, one problem arises: making changes on the mesh at the boundaries of each partition is very difficult. Every partition involved in the remeshing of the local patch of elements needs to perform exactly the same operations but without all the information required (as some part of the local patch of elements is not present in its memory). This has been addressed by using the properties of the repartitioning process as explained further.\\

In this section the parallel implementation of the TRM model will be explained, all situations described before will be addressed below. 

\subsection{Initial partitioning}\label{sec:InitialPartitionning}

\begin{figure}[!h]
\centering
\includegraphics[width=0.6\textwidth] {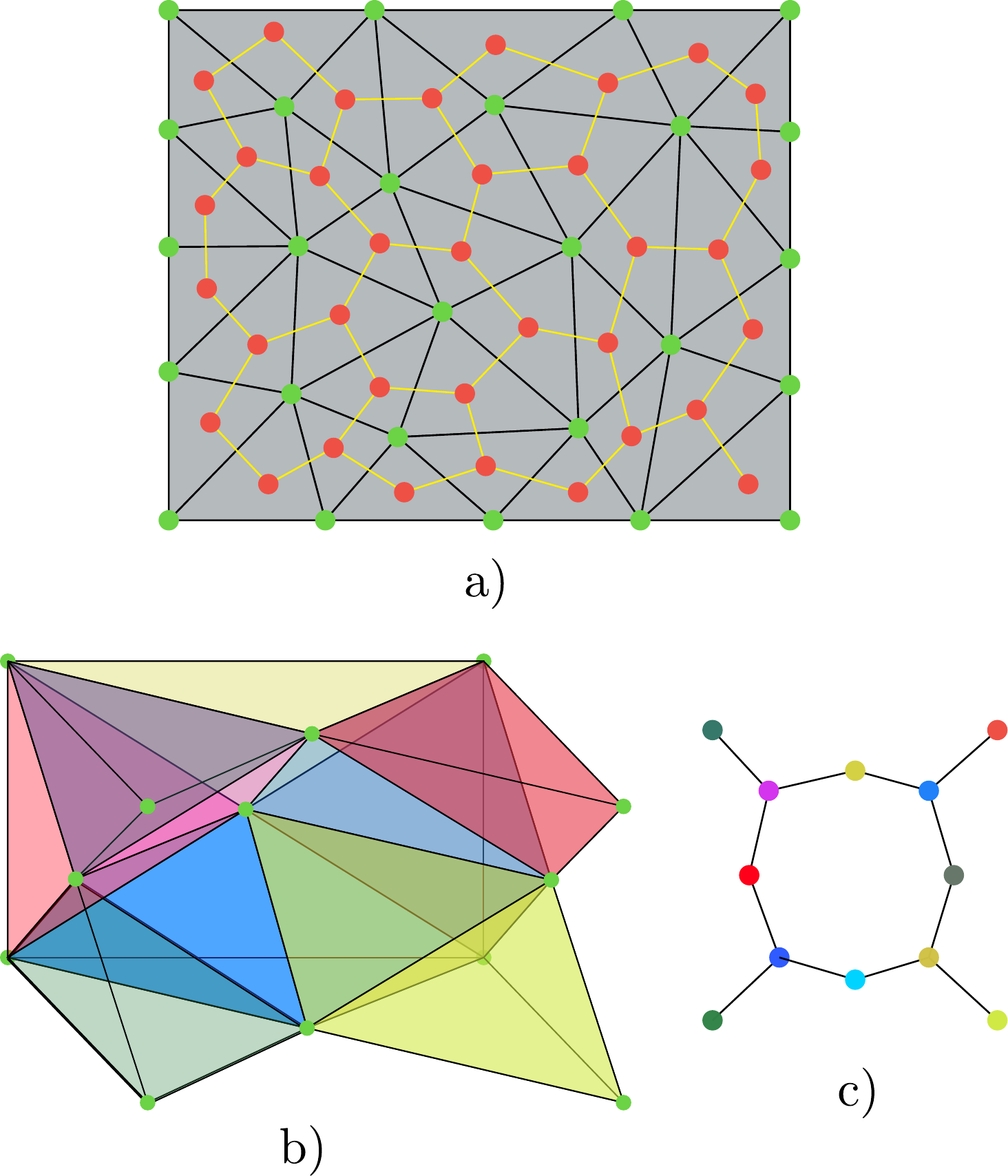}
\caption{Example of dual graph in 2D and 3D. a) mesh and dual graph of the mesh, green points are nodes from the mesh and red points are nodes from the dual graph of the mesh, b) mesh in 3D c) dual graph of the mesh in b).}
\label{fig:DualMesh}
\end{figure}

Two ways of partionning an Eulerian mesh are classically used: the first is to partition the graph of the mesh (each node is a vertex in the graph and each edge of the elements is an edge on the graph) to obtain multiple subsets of nodes. Each partition will receive a subset where their nodes are not present in any other subset. The second is to compute the partition over the dual graph of the mesh (or dual graph of a planar graph\footnote{a graph that can be projected into a plane and that its edges intersect only at their nodes}) of the initial mesh. In the dual graph, each vertex represents an element of the initial planar graph and each edge represents an edge of the initial planar graph that is shared by two elements. A planar graph may have multiple dual graphs, depending on the embeding of the planar graph in the plane (at this stage it is no longer a planar graph but a plane graph). However once this embedding is defined, the dual graph of the plane graph is completely defined. Coherent Eulerian meshes, are planar graphs: in a 2D context, the computation of the dual mesh uses each 2-simplex (elements) as a vertex and each 1-simplex (edges) as an edge, while in a 3D context each 3-simplex (elements) is treated as a vertex and each 2-simplex (facets) as an edge. Partitioning the dual graph of the mesh produces subsets of elements of the initial mesh, each partition will receive a subset where their elements are not present in any other subset. Figure \ref{fig:DualMesh} shows examples of the dual graph of a mesh in 2D and in 3D.\\

\begin{figure}[!h]
\centering
\includegraphics[width=0.7\textwidth] {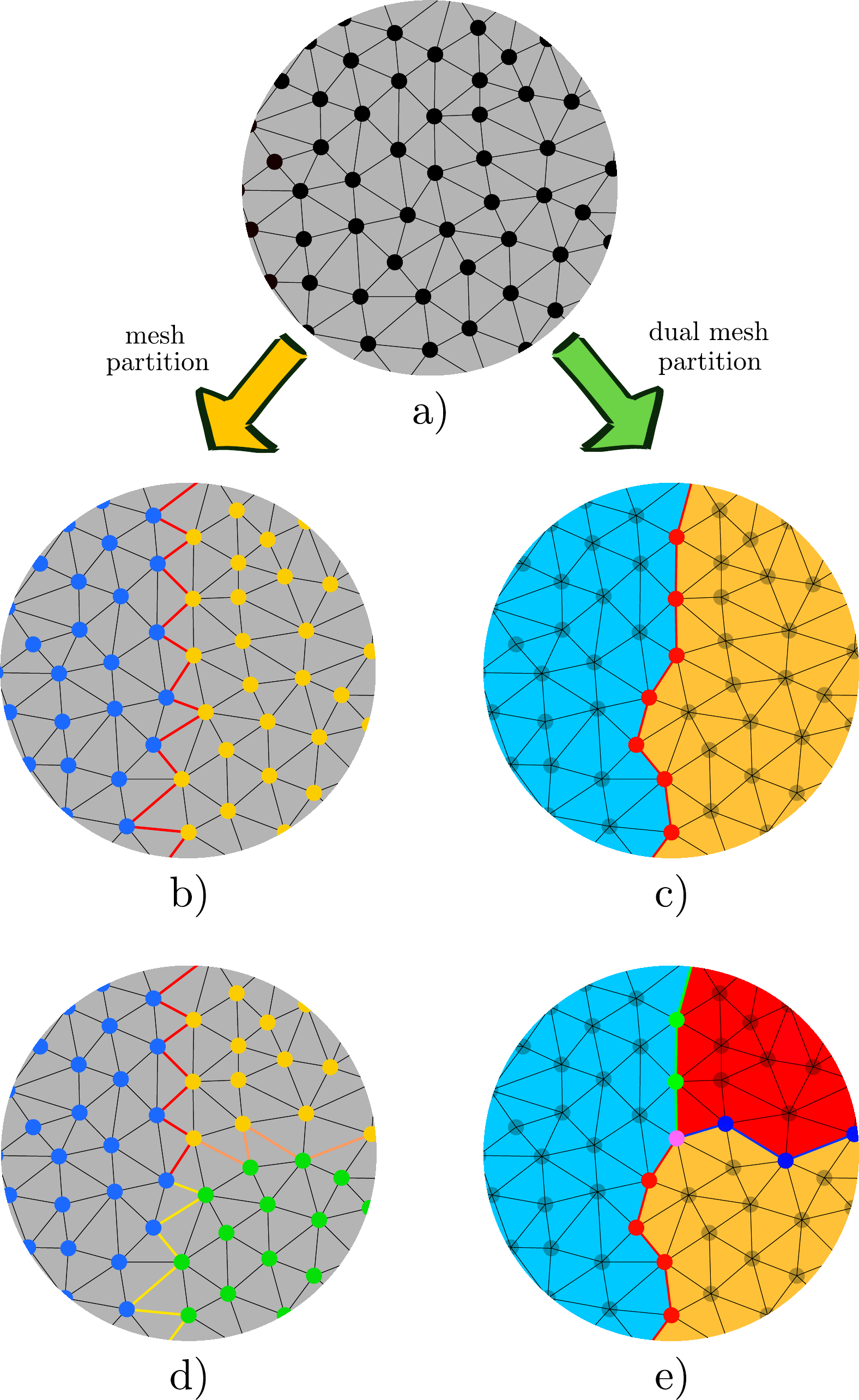}
\caption{Example of partitioning using the mesh and the dual mesh. a) mesh to partitionate, b) and d) partitioning using the mesh in two parts and three parts respectively, each part is represented by a color on the nodes, each part receives nodes not belonging to any other partition, some vertex (in color) are crossed by the partitioning. c) and e) partitioning using the dual mesh in two parts and three parts respectively, each part is represented by a color on the elements, each part receives elements not belonging to any other partition, some nodes (colored nodes) are in several partitions at the same time, these are Shared-Nodes. }
\label{fig:PartitionAssembly}
\end{figure}

Metis is able to make partitions using either the graph of the mesh or the dual graph of the mesh. In our context we have choosen to use the dual mesh. When a mesh is directly partitioned some edges of the mesh crossed by the boundaries of partitions must be managed. While it is still possible to use this configuration in a parallel context by repeating the elements that contain at least one edge crossed by the partitions as in \cite{Coupez2000, Mesri2009}, in our context where a certain geometrical data structure must be respected, using such a way of partitioning is more complex. By partitioning the dual mesh, even though each part receives elements not belonging to any other part, some nodes can be shared by several parts at the same time and from this point forward these are defined as Shared-Nodes. Figure \ref{fig:PartitionAssembly} shows examples of partitions made using the graph of the mesh and the dual graph of the mesh in two parts and in three parts.

Once the mesh is partitioned, each processor receives a part of the elements to work with that are more or less equally distributed among all the other processors. However this is true only for the preprocessor of the whole simulation, once the TRM model starts remeshing and changing the position of nodes, the charge of each processor evolves (see section \ref{sec:repartitionning} for this aspect).\\

\subsection{Numbering geometric entities and regularization} \label{sec:gluing}

\begin{figure}[!h]
\centering
\includegraphics[width=1.0\textwidth] {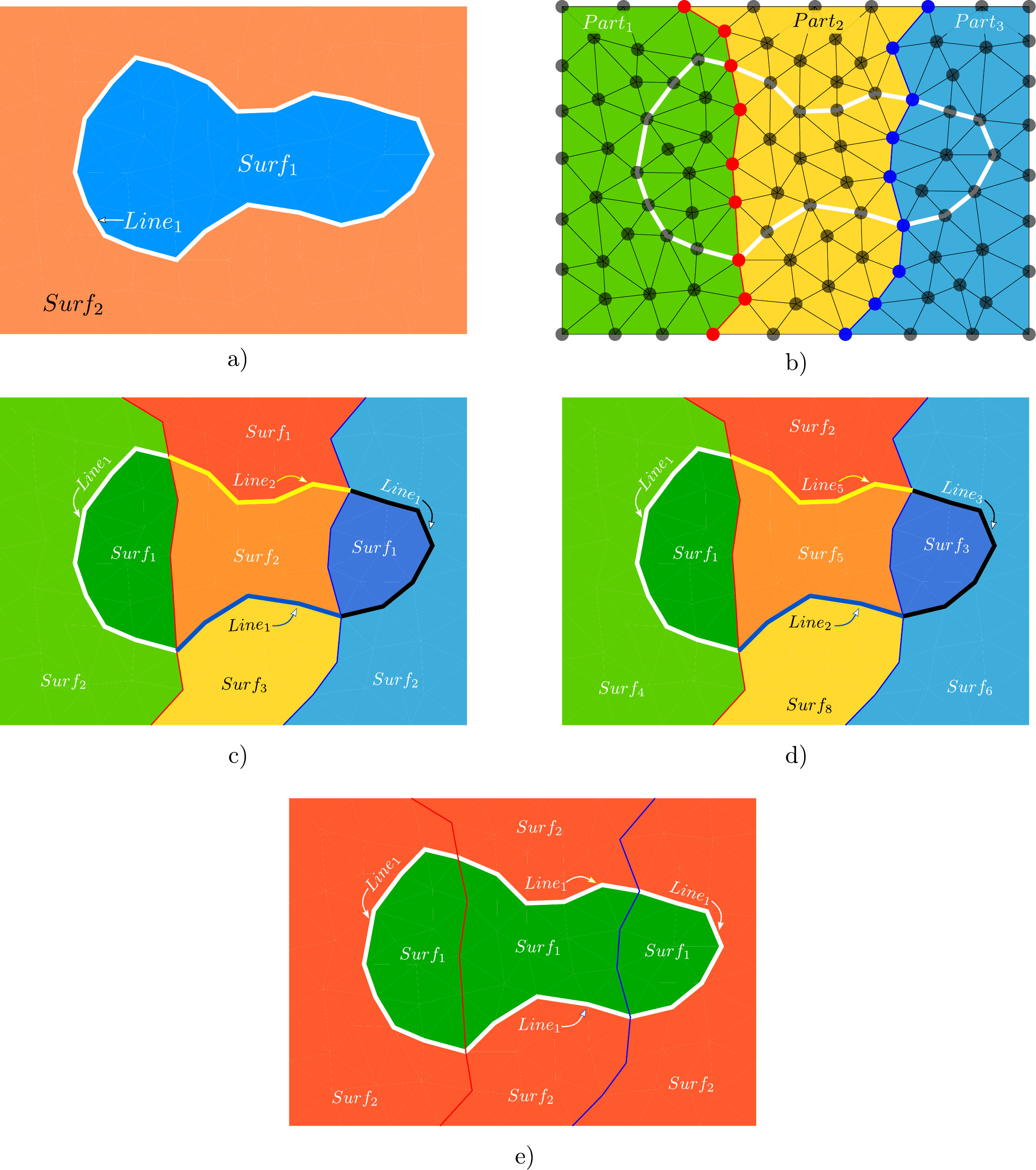}
\caption{Example of the initial numbering of geometric entities of a domain crossed by the boundaries partitions. a) geometric entities to number and also sequential numbering b) the same  domain partitioned in three parts, each color represents a partition, c) configuration of the geometric entities on each part of the domain, the numbering is done by default d) configuration of the geometric entities on each part of the domain, the numbering for each geometric type (Point, Line or Surface) begins with the number of the local partition and increases by the number of total partitions, e) configuration of the geometrical entities after renumbering.}
\label{fig:GluingEntities}
\end{figure}

In the present parallel context, it is possible that some gemetric entities appear in more than one partition at the same time. This creates issues as the algorithms to reconstruct the geometric entities had been developed in a sequential context (see section 2.2 of \cite{Florez2020b}). By performing the sequential algorithms for the reconstruction of entities, each partition will consider that the geometric entities stop at the boundaries between partitions. Problems as the one presented in figure \ref{fig:GluingEntities} could arise: in this configuration with two surfaces and a line, the expected identification of each geometrical entity should be the one illustrated in figure \ref{fig:GluingEntities} a), instead as each partition does not contain all the information the identification of entities would be as illustrated in figure \ref{fig:GluingEntities} c) where multiple geometric entities of the same type (Point, Line or Surface) can have the same identity (i.e. $Surf_1$ from $Part_1$ and $Surf_1$ from $Part_2$) but do not correspond to the same entity, or inversely, entities having different identities within the same partition but being the same entity in the whole domain (i.e. $Line_1$ and $Line_1$ from $Part_2$ or $Surf_1$ and $Surf_3$ from $Part_2$).\\

We have solved these situations in a two step process: first a non-repeating numbering system in parallel and then a regularization of the identity of entities crossed by the partitions.\\

The non-repeating numbering is a very simple process: Each geometric type (Point, Line or Surface) will be numbered according to the partition where it is present, starting with the number of the local partition (i.e. if numbering in $Part_3$ the first Line will be numbered as $Line_3$ the first point as $Point_3$ and the first surface as $Surf_3$) subsequently the numbering increases by the number of total partitions (i.e. if numbering in $Part_3$ and the total number of partitions is 4 the second Line will be numbered as $Line_7$). This numbering system is illustrated in figure \ref{fig:GluingEntities} d). This solves the problem that multiple entities can be identified as the same in different partitions but not correspond to the same (see Fig. \ref{fig:GluingEntities} c)).\\

The regularization helps to identify the entities that have been crossed by the boundaries of the partitions. This procedure is implemented via a local identification of the entities coupled to the shared-nodes, a communication process of that information, a remote identification of all entities and a re-numbering of the entities that need it. Algorithm \ref{alg:IdentityReg} contextualizes this procedure with the help of a recursive function presented in Algorithm \ref{alg:RecursiveTripletTreatment}. Note that the lines 4: and 10: of Algorithm \ref{alg:IdentityReg} send information to other partitions. These communication operations are realized in the actual implementation with the help of some predefined functions of MPI (MPI\_Alltoallv and MPI\_AllGatherv) performed over all processors. The purpose of the call of Algorithm \ref{alg:RecursiveTripletTreatment} in line 14: is to fill the set $SameEntity$. This set gathers list groups the identities given to a single geometric entity over all partitions, this list is computed with the same information in all processors hence it gives the same answer in all of them. In the configuration given in Figure \ref{fig:GluingEntities} d), when Algorithm \ref{alg:IdentityReg} is performed over the Surfaces (the geometric entities defining Surfaces in the TRM model), $SameEntity$ would store two lists one for each iteration of the while loop of line 11: in the first iteration the list would be composed of three pairs: $\{\{Surf_1, Part_1\},\{Surf_5, Part_2\},\{Surf_3, Part_3\}\}$ and the second iteration of four pairs: $\{\{Surf_4, Part_1\},\{Surf_8, Part_2\},\{Surf_2, Part_2\},\{Surf_6, Part_3\}\}$. On each iteration, the entity will be named on all partitions with the lowest identity found for it, hence in our example, $Surf_5$ and $Surf_3$ will be named $Surf_1$ in their respective partitions and $Surf_4$, $Surf_8$, and $Surf_6$ will be named $Surf_2$ in their respective partitions. This finally solves the inconsistencies found by the initial numbering of entities (see Fig. \ref{fig:GluingEntities} e)).\\

At this stage one final situation that needs solving appears: if there are non \emph{locally connected}\footnote{Geometric entities that are connected on partitions different from where part of them are stored} entities in the same partition with the same identity (e.g. $Surf_2$ in $Part_2$ and $Line_1$ in $Part_2$ of Fig. \ref{fig:GluingEntities} e)) these entities have to be gathered in one single entity. This procedure is made easily for surfaces as it is only necessary to exchange the nodes and elements from one Surface to the other and to remove the one that remains empty. On the other hand, in order to regroup Lines, an additional information is needed as one Line can now be made of disconnected segments and this was not taken into account in the algorithms developed in \cite{Florez2020b}. This has been solved by adding an additional information to L-Nodes: the \emph{previous} and the \emph{next} node on its Line. If there is no previous node or next node at that position in the Line, the value attributed to it is $Null$. i.e. If in a Line $L_1$ the L-Node $b$ is the first Node, the L-Node $c$ is the second node and the Line has an initial Point with a P-Node $a$, the previous and next nodes of $b$ are $a$ and $c$ respectively and the previous node of $c$ is $b$. Note that node $a$ does not have information about its previous and next node as it is a P-Node and not a L-Node. This implementation solves the fact that one line can be segmented in multiple parts without need of further changes to the already existent code.

\begin{algorithm}
\caption{Identity Regularization Algorithm performed on $Part_i$}\label{alg:IdentityReg}
\begin{algorithmic}[1]

\ForAll{Shared-Nodes : $N_i$}
	\ForAll{SharedRanks of $N_i$ : $Part_j$}
		\State $I_{L}$ $\gets$ local identity of the coupled entity of $N_i$
		\State send to $Part_j$: Pair($N_i$, $I_L$)\footnote{A Pair is a structure of data with two objects, the function $First(Pair_i)$ extracts the first object in $Pair_i$ and the function $Second(Pair_i)$ extracts the second object}
	\EndFor
\EndFor

\ForAll{Parts: $Part_j \neq Part_i$}
	\ForAll{ Received Pairs from $Part_j$: $Pair_k$}
		\State $N_i$ $\gets$ the Node $First(Pair_k)$
		\State $I_L$ $\gets$ local identity of the coupled entity of $N_i$
		\State $I_R$ $\gets$ Second($Pair_k$)
		\State send to all Parts: Triplet($I_L$, $I_R$, $Part_j$)\footnote{A Triplet is a structure of data with three objects, the function $First(Pair_i)$ extracts the first object in $Pair_i$, the function $Second(Pair_i)$ extracts its second object and the function $Third(Pair_i)$ extracts its third object}
	\EndFor
\EndFor

\While{Triplets to Treat}
	\State $TT$ $\gets$ Take First Non-Treated Triplet

	\State Create empty list of Pairs: $SameEntity$ \Comment{[Identity, Partition]}
	
	\State Call \textbf{RecursiveTripletTreatment}($TT$, $SameEntity$)
	
	\State $I_{Lowest}$ $\gets$ the lowest value of the first item in all pairs of $SameEntity$
	\ForAll{Pairs in $SameEntity$ : $Pair_k$}
		\If{$Second(Pair_k) == Part_i$}
			\State $I_{Old}$ $\gets$ $First(Pair_k)$
			\State change identity of entity with number $I_{Old}$ to $I_{Lowest}$
		\EndIf
	\EndFor
\EndWhile

\end{algorithmic}
\end{algorithm}

\begin{algorithm}
\caption{Recursive function for the identification of the same entity given a triplet $Tx$ and a list of pairs to fill $P_L$}\label{alg:RecursiveTripletTreatment}
\begin{algorithmic}[1]
\Function {RecursiveTripletTreatment}{$Tx$, $P_L$}
	\State $I_L$ $\gets$ $First(Tx)$ (local identity)
	\State $I_R$ $\gets$ $Second(Tx)$ (remote identity)
	\State $Part_r$ $\gets$ $Third(Tx)$ (remote Partition)
	\ForAll{ Received Triplets from Part $RPart$ : $Triplet_i$}
		\If {$Triplet_i$ is still not treated and $First(Triplet_i)==I_R$}
			\State Set $Triplet_i$ as treated
			\State Add to $P_L$ : Pair($Second(Triplet_i)$, $Third(Triplet_i)$)
			\State Call \textbf{RecursiveTripletTreatment}($Triplet_i$, $P_L$)
		\EndIf
	\EndFor
\EndFunction
\end{algorithmic}
\end{algorithm}

\subsection{Re-Equilibrium of Charges, Repartitionning: Mesh Scattering}\label{sec:repartitionning}

Repartitioning in a distributed memory framework is much more complex than making an initial partitioning as all the information is scattered through all processes. Even though there exist tools that solve these kinds of problems (e.g. an extension of Metis called ParMetis \cite{Karypis1998}) we have opted to develop our own repartitionner. This is mainly because in the present context we have to make sure that the data structure of the different geometric entities stay consistent during the repartitioning process. Of course our objective is not to develop a partitioner having all the capabilities of the well established libraries of the domain, but to develop a simple and robust way of exchanging information between process having parts of an scattered Eulerian mesh on its memory.

\subsubsection{Dynamic Ranking System} \label{sec:DynamicRank}

The repartitionning algorithm developed for the TRM model will use a ranking system to determine the direction of the information flow, each process having its own unique rank\footnote{in our scope, these ranks refer to different ranks than those attributed to each process by MPI, at the beginning of the parallel environment.}. This rank can be determinated for example by using the number of elements, the number of nodes, the number of edges, or the total surface of the domain stored on each processes, and in the case were two process have the same value, a random attribution is held. For the purpose of this publication we will use the number of elements to obtain the rank $R_i$ of process $i$. Once each process has determinated the number of elements present on its memory, this information is sent to all other processors via MPI, each processor makes the comparison procedure and stores a list $RankingOrder$ of the ranks of each processor, the lower the number of elements on a partition, the higher is its rank. This list is computed equally on all processors. Moreover, if each part $i$ needs the rank of part $j$, this information is obtained via $RankingOrder[j]$.

\subsubsection{Shared-Nodes}

As defined before, Share-Nodes are the nodes belonging to multiple process at the same time, these nodes reside at the boundaries between partitions. Additionally to the Dynamic Ranking System, some other information will be required for the repartitioning operation, this information denotes for all Shared-Nodes, to which partitions they are shared. by considering for example figure \ref{fig:PartitionAssembly}(c), here the red nodes are shared by the cyan and by the yellow partitions, the red nodes from the cyan partitions must know that they are shared by partition blue and vice versa. Some nodes can be shared by more than two partitions as in Figure \ref{fig:PartitionAssembly}(e). In this particular case, three partitions share the magenta node. when this node is processed the blue part must know that it is shared by the yellow part and the red part. This information can be obtained very easily: each part communicates via MPI to all processes which nodes are at its boundary, then for all received nodes, if the node exist in the current part, it means it is shared with the sender part. Every node then stores a list $SharedRanks$ with the parts containing it.\\

\subsubsection{Unidirectional Element Sending} \label{sec:ElementSending}

As explained in section \ref{sec:InitialPartitionning}, our parallel scheme maintains sets of non-repeated elements scattered in all processes, this means also that the repartitioning scheme must exchange elements between partitions maintaining also this property. If a partition sends one element, this element must have only one destination to maintain coherence in the proposed algorithm. This goal is achieved using the ranking system and the $SharedRanks$ list of the Shares-Nodes in Algorithm \ref{alg:UnidirEleSel}.\\

\begin{algorithm}
\caption{Unidirectional element selection in Part $Part_i$}\label{alg:UnidirEleSel}
\begin{algorithmic}[1]

\State Store a list of list: SharedNodesPerPart
\ForAll{Shared-Nodes : $N_i$}
	\ForAll{SharedRanks of $N_i$ : $Part_j$}
		add item $N_i$ to SharedNodesPerPart[$Part_j$]
	\EndFor
\EndFor

\State Store a list of list: ElementsToSendToPart

\ForAll{Parts: $Part_j$}
	\If{RankingOrder[$Part_i$]$<$RankingOrder[$Part_j$]}
		\ForAll{ SharedNodesPerPart[$Part_j$] : $N_j$}
			add $Elements(N_j)$\footnote{Function $Elements(Entity)$ extracts the Elements present in $Entity$, where $Entity$ can be a surface or a node. When extracting the elements from a node $N$, the result will be the set of elements surrounding $N$.} to ElementsToSendToPart[$Part_j$]
		\EndFor
		\State take out repeated elements in ElementsToSendToPart[$Part_j$]
	\EndIf
\EndFor

\ForAll{Parts: $Part_j$}
	\ForAll{ ElementsToSendToPart[$Part_j$] : $E_j$}
		\ForAll{Nodes($E_j$)\footnote{Function $Nodes(Entity)$ extracts the nodes present in $Entity$, where $Entity$ can be a point, a line, a surface or an element.}  : $N_k$}
			\ForAll{SharedRanks of $N_k$ : $Part_k$}
				\If{RankingOrder[$Part_j$]$<$RankingOrder[$Part_k$]}
					\State Erase $E_j$ from ElementsToSendToPart[$Part_j$]
				\EndIf
			\EndFor
		\EndFor
	\EndFor
\EndFor

\end{algorithmic}
\end{algorithm}

In the first part of this Algorithm \ref{alg:UnidirEleSel}, two lists are created, these lists contain the Share-Nodes and the Elements to send to each partition. Note that each processor performs this algorithm, hence, each processor has a list of elements to send to all the other processors (even if that list is empty). A first selection is made via the inequality of line 6 of Algorithm \ref{alg:UnidirEleSel}, where the list of elements to export only accepts the elements that have at least one node shared with a partitions with a superior rank (a partition with a lower number of elements). In the last section of the algorithm, a final selection is made for the elements that are in the list to be sent to multiple processors (i.e. for elements with nodes shared by more than two processors at the same time). The final destination selection for these contentious elements is achieved by choosing the partition with the highest rank (lines 10: to 15: of Algorithm \ref{alg:UnidirEleSel}), erasing the element in question from all the other lists. Once this algorithm is executed in all partitions, a \emph{scattering} process begins, sending all elements to their new partitions along with some associated information: Node positions, Node fields, element fields and some data necessary to fill the data structure of the geometric entities involved in the scattering (see section \ref{sec:GeoReconstruction}).\\

\begin{figure}[!h]
\centering
\includegraphics[width=0.5\textwidth] {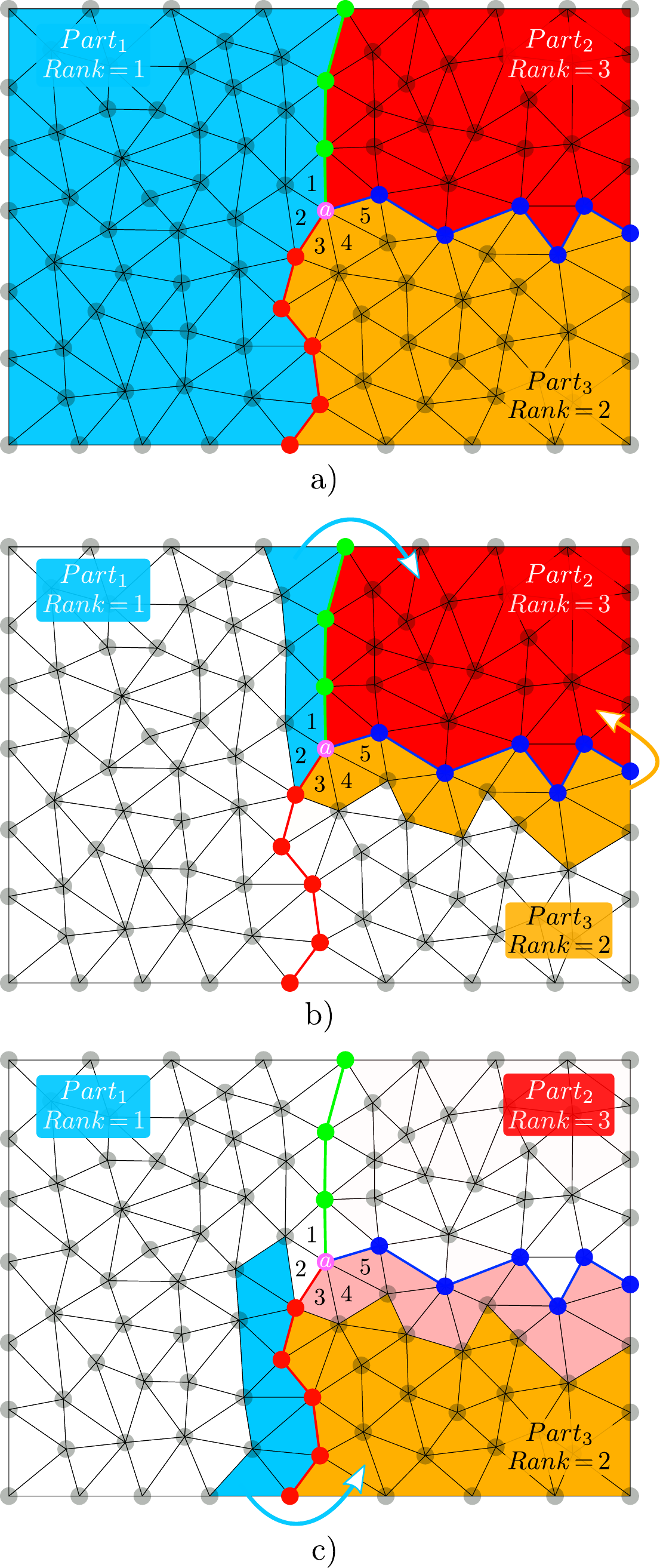}
\caption{ Example of the behaviour of the Unidirectional selection element algorithm. a) initial state with three parts, the name and the rank of each partition is displayed. b) Selected elements to be sent to $Part_2$, elements 3, 4 and 5 appear only to be sent to $Part_2$, and not to $Part_1$. Elements 1 and 2 appear initially on the list to be sent to $Part_2$ and $Part_3$ but they are filtered in the last part of the algorithm, as the higher rank of the nodes of these elements belongs to $Part_2$. c) selected elements to be sent to $Part_3$, the intersection of elements from $Part_1$ to be sent to $Part_2$ and $Part_3$ is empty.}
\label{fig:UnidirectionalSelection}
\end{figure}

\begin{figure}[!h]
\centering
\includegraphics[width=0.55\textwidth] {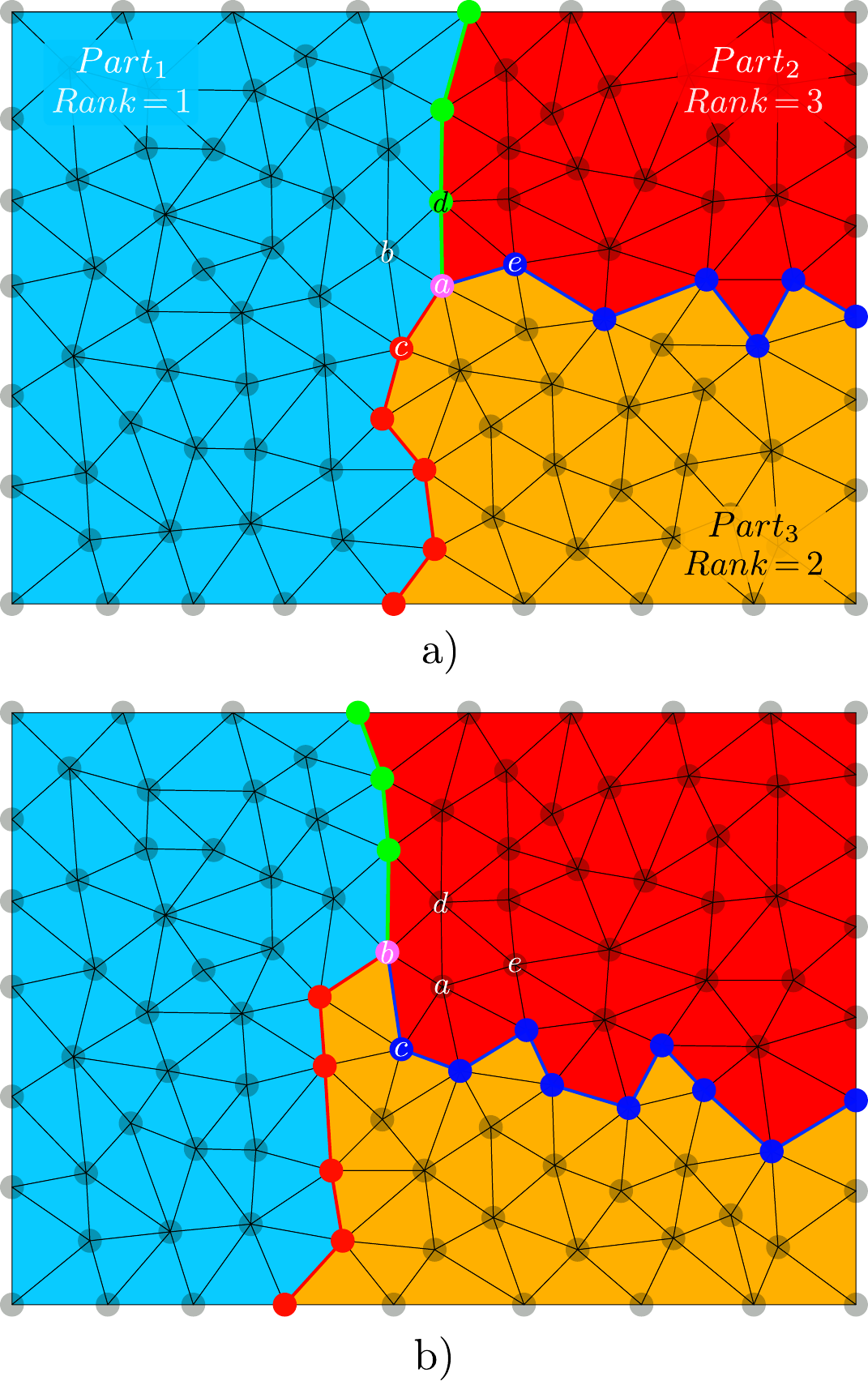}
\caption{Example of the new configuration of each part after the scattering of figure \ref{fig:UnidirectionalSelection} to their respective new partitions, the boundaries of all parts are displaced by the scattering including the shared node between the three parts a) initial configuration, node a is shared by the three parts, b) configuration after scattering, now node b is shared by the three parts while node a is part of the bulk nodes of part 2.}
\label{fig:UnidirectionalScattering}
\end{figure}

Figure \ref{fig:UnidirectionalSelection} illustrates one example of the behaviour of the unidirectional selection element algorithm for three partitions. The rank of each partition is computed according to the number of elements in decreasing order. In Fig. \ref{fig:UnidirectionalSelection} b) it is shown the elements to be sent to $Part_2$. Here, Elements $3$, $4$ and $5$ appear only to be sent to $Part_2$, and not to $Part_1$. This filter is applied at the initial part of the algorithm as the inequality RankingOrder[$Part_3$]$<$RankingOrder[$Part_1$] is not evaluated to true. Elements $1$ and $2$ appear initially on the list to be sent to $Part_2$ and $Part_3$ but are filtered in the last part of the algorithm, as $Part_2$ is of higher rank. In Fig. \ref{fig:UnidirectionalSelection} c) the selected elements to be sent to $Part_3$ are displayed. Note that some of the elements of $Part_3$ are going to be sent to $Part_2$ hence they appear in a different color. Also, the intersection of elements from $Part_1$ to be sent to $Part_2$ and $Part_3$ is empty, hence no errors will be made in the scattering process. Figure \ref{fig:UnidirectionalScattering} illustrates the initial and the final configuration after the scattering. The boundaries of all parts are displaced by the scattering including the shared node between the three parts (Node $a$) after the scattering, Node $b$ is shared by the three parts while Node $a$ is part of the bulk nodes of $Part_2$.\\

\subsection{Geometry reconstruction} \label{sec:GeoReconstruction}

In the parallel re-equilibrium of charges explained in section \ref{sec:repartitionning}, a particular problem appears. In fact, when exchanging elements and nodes between partitions, a reconstruction of the geometrical entities that involve those elements and nodes have to be considered. Having the data structure presented in section 2.1 of \cite{Florez2020b}, it is clear that these reconstructions must be addressed depending on the type of geometry in question. While the case of Surfaces is trivial (they move along with elements), the consideration of the geometric entities attached to L-Nodes and P-Nodes is more complex.

\begin{figure}[!h]
\centering
\includegraphics[width=0.8\textwidth] {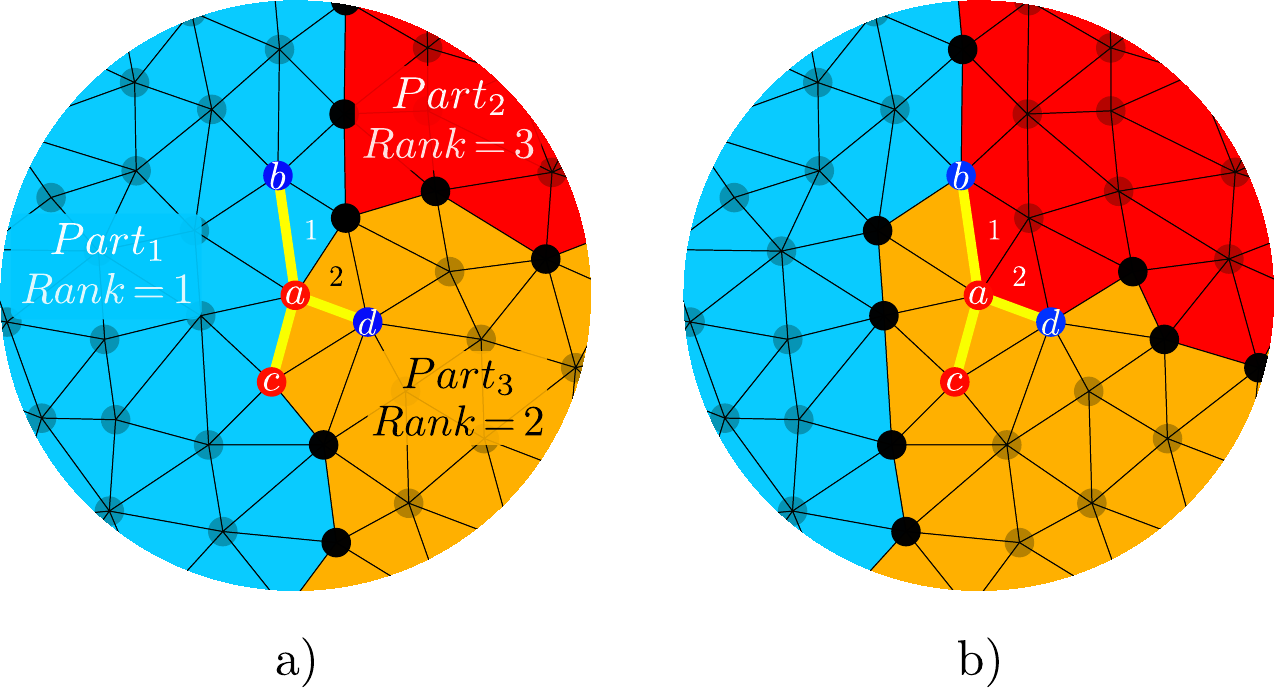}
\caption{Example of the scattering near a P-Node $a$ the corresponding connections of its coupled Point are displayed in yellow to L-Nodes $b$ and $d$ and to P-Node $c$. a) Initial configuration, connections to the Point of P-Node $a$ are stored in $Part_1$ and $Part_3$ and b) Configuration after the scattering, connections to the Point of P-Node $a$ are distributed within all parts}
\label{fig:PointReconstructionTriple}
\end{figure}

When reconstructing the entities coupled to the new L-Nodes and P-Nodes (Lines and Points) more information is needed from the sending part: new P-Nodes on the receiving partition need to create new Points, hence, information regarding the connections of that point to other points and lines is needed (see the data structure of points of section 2.1 of \cite{Florez2020b}). In some cases, this information needs to be collected from multiple partitions at the same time as illustrated in Figure \ref{fig:PointReconstructionTriple}. Here the scattering procedure is being performed near a Point attached to the P-Node $a$, in the initial configuration (Fig. \ref{fig:PointReconstructionTriple} a)). This Point is in the memory of $Part_1$ and $Part_3$ (hence P-Node $a$ is a Shared-Node). The connections of L-Nodes $a$ and $b$ are only known by $Part_1$ and $Part_3$ while the connection to P-Node $c$ is known by both partitions, note that $Part_2$ does not have any of the previous information regarding the connections of P-Node $a$. After the scattering, elements 1 and 2 were sent to $Part_2$ by $Part_1$ and $Part_3$ respectively, thus both partitions need to send the adjacent information to create all the nodes unkown by $Part_2$. $Part_1$ sends the node $a$ and $b$ and $Part_3$ sends the node $a$ and $d$, both partitions sent information regarding the node $a$ because they have no way to know that the other partition has already sent it. This is not necessarily a costly step because the information regarding the connections of the Point of P-Node $a$ can be attached to the communication. Indeed, information regarding all connections of the Points coupled with the P-Nodes sent to other partitions is also broadcasted. This means that $Part_1$ sends the P-Node $a$ along with the identities $c$ and $b$  and $Part_3$ sends also the P-Node $a$ along with the identities $c$ and $d$ corresponding to the connections known by $Part_1$ and $Part_3$ respectively. $Part_2$ receives two times the P-Node $a$, however it is only created once. On the other hand, the information received about its connections is used: $Part_2$ searches on its memory for the existence of the nodes $b$, $c$ and $d$ received in the communication to be connected to the Point of P-Node $a$ and if they exist, the connection is made. Note that the P-Node $c$ can not be found by $Part_2$ hence that connection can not be created.\\

\begin{figure}[!h]
\centering
\includegraphics[width=0.96\textwidth] {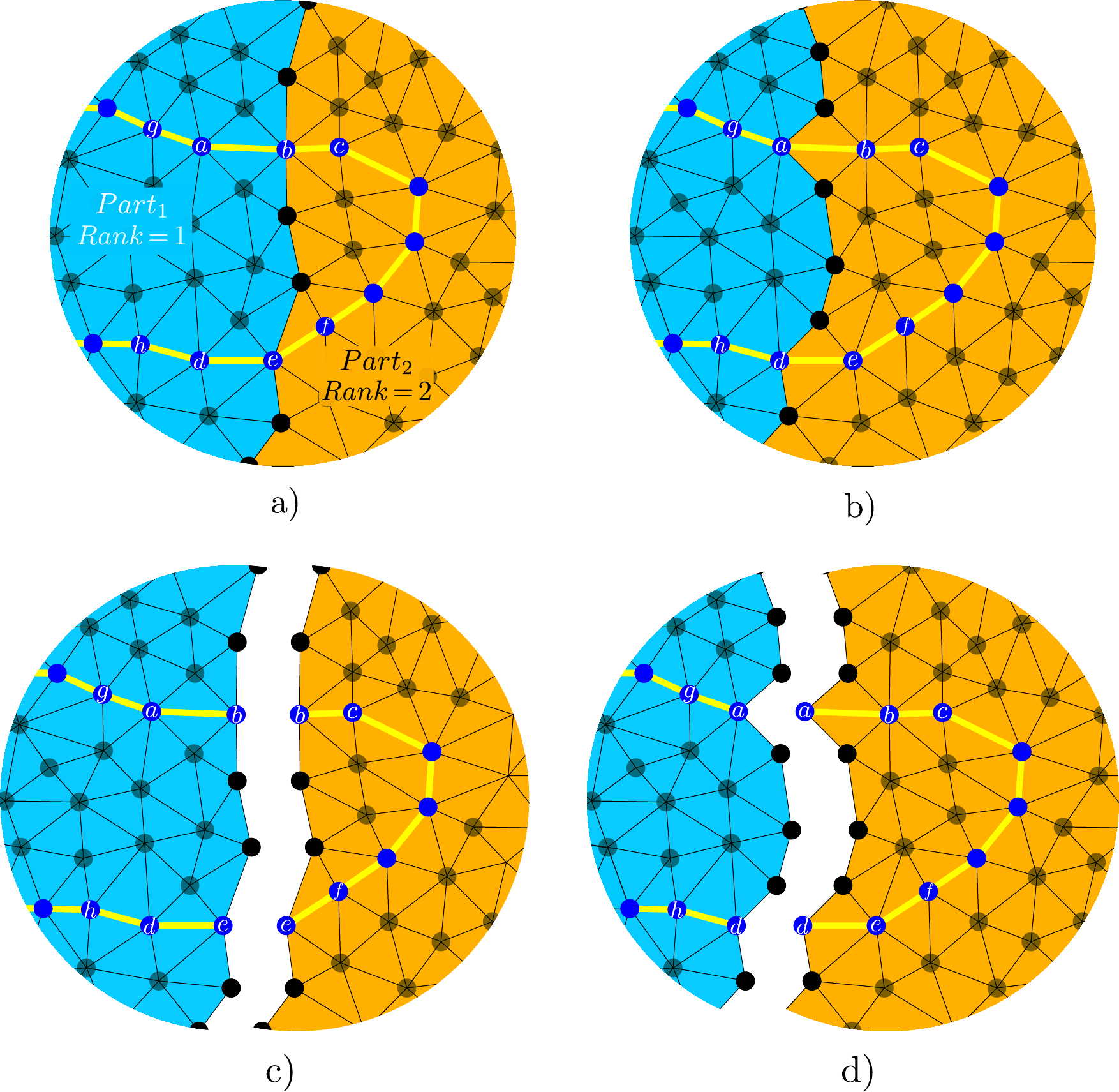}
\caption{Example of the scattering with a Line (yellow) crossed by the boundary of two partitions $Part_1$ and $Part_2$, the assembled and separated views are displayed, the assembled view shows the domain being simulated while the separated view show the actual memory of each partition. a) and c) Initial configuration, assembled and separated views respectively, b) and d) configuration after scattering, assembled and separated views respectively.}
\label{fig:LineReconstructionDouble}
\end{figure}

Line reconstructions are performed in a similar manner as for Points. The data structure of Lines is special as it is composed of an ordered sequence of L-Nodes and of optional initial and final Points as stated in section 2.1 of \cite{Florez2020b}. Furthermore, in section \ref{sec:gluing} of this article an additional property was included for L-Nodes: information regarding the previous and next nodes within the same Line was added. This additional property helps to reconstruct the Lines involved in the scattering by sending the identity of the previous and next nodes along with the L-Node, additionally to the identity of the Line where it must be added. The position of the L-Node within that Line is then obtained by its relative position to these next and previous nodes if they exist in the memory of the partition (similarly to the reconstruction of Points). If one of those nodes do not exist, it means that the Line is segmented at that L-Node. Figure \ref{fig:LineReconstructionDouble} shows the initial and final states of a scattering between two partitions where a Line has been crossed by the boundary of two partitions $Part_1$ and $Part_2$. In the scattering, two L-Nodes are sent by $Part_1$ to $Part_2$. Node $a$ carries the identity of its previous (L-Node $g$) and next (L-Node $b$) nodes. Similarly, node $d$ carries the identity of its previous (L-Node $e$) and its next (L-Node $h$) nodes. This information helps to connect the new nodes in $Part_2$ to the Line in their respective positions and to fill the information about the previous node of L-Node $b$ and the next node of L-Node $e$ (and vice versa). Note that even though the information about the previous node of L-Node $a$ was sent, this information is not used as the node $g$ does not appear in the memory of $Part_2$ (see Fig. \ref{fig:LineReconstructionDouble} d)). Consequently, the information of the previous node of L-Node $a$ is sey as $Null$.\\

\subsection{Blocking remeshing at partition boundaries} \label{sec:blockedOperations}

Making changes to the mesh at the boundaries of each partition is very difficult because part of the local patch of elements is not present in its memory. Every partition involved in the remeshing of the local patch of elements needs to perform the same operation but without all the information required. We have solved this in a simple manner by blocking all operations involving a change (change of connectivity or of position) on the edges belonging to the boundaries between partitions. For example, in figure \ref{fig:UnidirectionalScattering} a), Node Collapse is blocked between nodes $a$ and $c$, Edge Splitting and Edge Swapping is also blocked for the edge $\overline{ac}$ delimited by the same nodes. However, note that this is not the case for nodes $d$ and $e$, while Node Collapse is still blocked between those nodes, Edge Swapping and Edge Splitting is allowed for edge $\overline{de}$, as this would not change the configuration of the edges of the boundaries between partitions. Other operations are also blocked as the Vertex Smoothing and the Node Glidding operations which are not allowed for any of the nodes at the boundary between partitions.\\

The repartitioning approach was also developed as a solution for the complexity of the remeshing process in parallel. Indeed, the repartitioning process presented in section \ref{sec:repartitionning} exchanges a complete layer of elements between the sending partition and the receiving partition, hence completely changing the boundaries every time it is performed. Figure \ref{fig:UnidirectionalScattering} b) illustrates how all the blocked operations mentioned above between edges $\overline{ac}$ and $\overline{de}$ are unblocked since these edges no longer belong to the boundary. A new remeshing process can be executed for all elements and nodes involving the previously blocked (now unblocked) edges to complete the remeshing process of the whole domain. The solution for the remeshing process is not the same as if it would be performed in a sequential framework as the different remeshing operations are not performed in the same order. However the influence of this artifact is expected to be minimal. This will be proven and discussed further in sections \ref{sec:results} and \ref{sec:conclusions}.\\

\subsection{other parallel treatments}
At this stage, two particular problems need to be solved. These particular situations correspond more directly to the physical model. However they may appear for the majority of boundary migration problems modeled with the TRM model. The first, is the computation of properties at Shared-Nodes and the second is the Lagrangian movement of these nodes. Within the parallel framework used for the TRM model, Shared-Nodes do not have all the necessary information of their sourrondings in the total domain, hence the treatments mentioned above can not be computed in the default sequential way as in \cite{Florez2020b}.\\

\subsubsection{Computation of properties at Shared-Nodes} \label{sec:PropSharedNodes}

\begin{figure}[!h]
\centering
\includegraphics[width=0.85\textwidth] {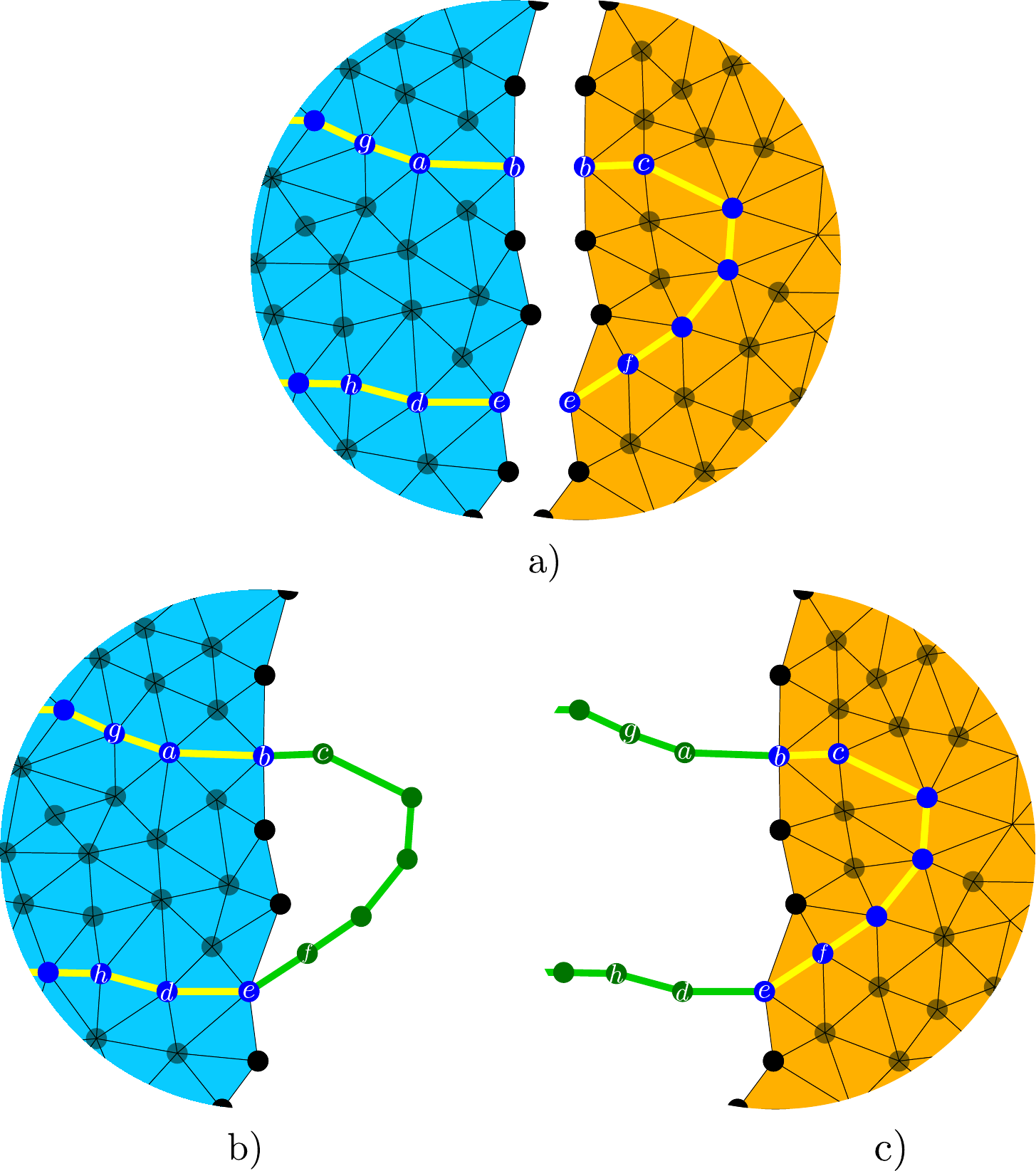}
\caption{Example of Line patch completion with temporary nodes (nodes in green), a) configuration for two partitions, b) $Part_1$ after adding the needed temporary nodes c) $Part_2$ after adding the needed temporary nodes.}
\label{fig:LinePatchCompletion}
\end{figure}

For GG mechanism, the necessary geometric properties to be computed are the mean curvature $\kappa$ and the normal $\vec{n}$  of the interfaces. As mentioned in section \ref{sec:theTRMmodel}, these properties are computed for L-Nodes using piecewise polynomials of third order (Natural Parametric Splines \cite{DeBoor1978}). Hence the information needed is the position of the nodes contiguous to the Shared-Nodes of each partition. This information is transmitted by the neighbouring partitions and stored in \emph{temporary} nodes. Figure \ref{fig:LinePatchCompletion} shows this situation, here the blue nodes correspond to L-Nodes belonging to the partition while the green nodes correspond to the temporary nodes obtained. Geometrical properties can now be computed for the L-Nodes $b$ and $e$ on each partition, obtaining the same result in both of them. These temporary nodes only exist during the computation of the geometrical properties. A similar operation is done for P-Nodes however only one contiguous temporary node is necessary for the computation of model II of \cite{Kawasaki1989} (used for the computation of the points's velocity in our model).\\

\subsubsection{Lagrangian movement in Parallel} \label{sec:LagrangianMovPar}

\begin{figure}[!h]
\centering
\includegraphics[width=0.7\textwidth] {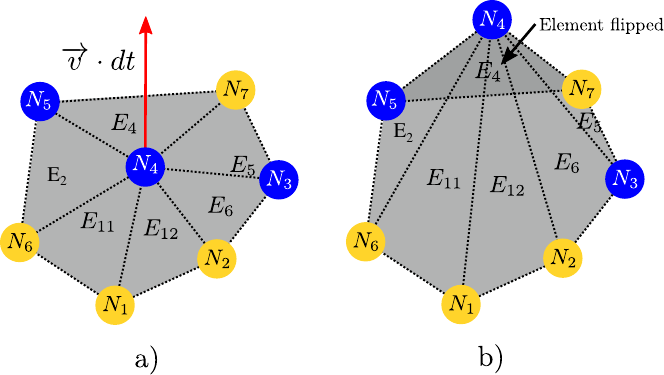}
\caption{Example of element flipping, a) initial state, the displacement vector of $N_4$ $\vec{v} \cdot dt$ lies outside the element patch. b) state after updating the position of $N_4$, element $E_4$ has been flipped. \cite{Florez2020b}}
\label{fig:FlippingExample}
\end{figure}

The Lagrangian movement presents a very similar problem as the one exposed in the previous section. In section \ref{sec:theTRMmodel}, it is shown that the Lagrangian movement checks for invalid configurations every time one nodes moves (see Fig. \ref{fig:FlippingExample}). In our parallel context this can not be checked for all the elements when a Shared Node is moved as some of these elements belong to other partition and do not exist on the memory of the local partition. The solution to this has been implemented as follows: considering that all Shared-Nodes computed the same velocity (using the method of section \ref{sec:PropSharedNodes}), each partition makes their respective flipping checking for the elements known to them. If a flipping is encountered, this is communicated to all partitions where the Node is shared and the movement is stored for retrying with half the displacement in a further iteration, thus obtaining the same response as it would be in a sequential implementation.\\

\subsection{The TRM algorithm for Grain Growth in parallel}

The final algorithm for a time step of the TRM model in the context on GG is summarized in Algorithm \ref{alg:TRMGGPar}.

\begin{algorithm}
\caption{Grain Growth TRM Algorithm in parallel}\label{alg:TRMGGPar}
\begin{algorithmic}[1]
\State \textbf{Remeshing Algorithm over non-blocked entities} (Algorithm \ref{alg:Remeshing} with the constraints explained in sec \ref{sec:blockedOperations})
\State \textbf{Recompute Ranking System} (section \ref{sec:DynamicRank})
\State \textbf{Mesh Scattering} (section \ref{sec:ElementSending})
\State \textbf{Reconstruct Geometries} (section \ref{sec:GeoReconstruction})
\State \textbf{Remeshing Algorithm over non-blocked entities}

\State \textbf{Complete Lines and Point Connections} (temporary nodes) (section \ref{sec:PropSharedNodes})

\ForAll{Points: $P_i$}
	\While{Number of Connections $>3$ }
		\State split multiple point $P_i$.
	\EndWhile
\EndFor
\ForAll{Lines : $L_i$}
    \State Compute the natural spline approximation of $L_i$.
\EndFor
\ForAll{L-Nodes : $LN_i$}
    \State Compute curvature and normal ($\kappa \vec{n}$) over $LN_i$.
\EndFor
\ForAll{P-Nodes : $PN_i$}
    \State Compute the product $\kappa \vec{n}$ over $PN_i$ using model II of \cite{Kawasaki1989}.
\EndFor

\State \textbf{Delete temporary nodes}

\ForAll{L-Nodes and P-Nodes : $LPN_i$}
	\State Compute velocity $\vec{v_i}$ of Node $LPN_i$
    
\EndFor

\State \textbf{Iterative movement with flipping check in parallel} (section \ref{sec:LagrangianMovPar})

\end{algorithmic}
\end{algorithm}

\section{Numerical results} \label{sec:results}

In \cite{Florez2020b} three test cases corresponding to the Circle-Shrinkage, T-Junction, and Square-Shrinkage tests, were performed in order to validate the accuracy of the TRM model. A final test was performed to validate the model in the context of a more realistic environment, where a 2D 10000 grain test was performed in order to compare the results to a classical LS-FE approach. Here we will perform 2 sets of 2D simulations over a large quantity of grains, using a strong and weak scaling benchmark: the first will consider a constant domain size performed with $N_p$ number of processors, this simulation will be denoted to be as ``Variable charge" (strong scaling) as the mean charge per processor will evolve inversely proportional to the number of processors. Then, a set of simulations at ``Constant Charge" (weak scaling) will be performed for a simulated domain increasing its size proportionally to the number of processors $N_p$ considered. For example, the simulation performed over 2 processors will consider a total domain surface twice as large as the one performed over 1 processor.\\

All simulations will be performed on a cluster facility composed of \emph{nodes}\footnote{here the meaning of the word node represent a physical CPU connected in the network of the HPC} Bullx R424 equiped with a chipset Intel Xeon E5-2680v4 with 28 processors at 2.4 GHz. The nodes are connected between them using an Infiniband FDR at a speed of 56 $Gb/s$. Hereafter we will use the syntax $N_n$x$N_p$ to describe the configuration used in our computations, where the term $N_n$ describes the number of CPUs used and $N_p$ the number of processors used per CPU, hence, for example for a simulation using 56 processors distributed on 2 CPUs the syntax 2x28 will be used.\\

\subsection{ Variable Processor Charge (strong scaling benchmark)}\label{VariableCharge}

Here two similar tests to the one presented in \cite{Florez2020b} with 10000 grains will be performed, extending the size of the domain to fit first 50000 and then 560000 grains, maintaining the initial grain size distribution, the same thermal treatment (1 hour at $1050$ $^{\circ}C$) and identical physical properties: the generation of the initial tessellation will be performed with a Laguerre-Voronoi cell generation procedure \cite{imai1985voronoi,Hitti2012,Ilin2016} over a squared domain and the values for $M$ and $\gamma$ are chosen as representative of a 304L stainless steel at $1050$ $^{\circ}C$  (with $M=M_0*e^{-Q/RT}$ where $M_0$ is a constant $M_0=1.56\cdot 10^{11}$ $mm^4/Js$, $Q$ is the thermal activation energy $Q=2.8\cdot 10^5$ $J/mol$, $R$ is the ideal gas constant, $T$ is the absolute temperature $T=1323$ $K$ and $\gamma=6\cdot 10^{-7}$ $J/mm^2$) \cite{Scholtes2016, Maire2017}. Moreover the initial grain size distribution is imposed as a Log-Normal distribution curve with a median value of 0.017 $mm$ and a standard deviation of 0.006 $mm$. Additionally, maximal and minimal limits for the size of grains introduced are defined as 0.04 $mm$ and 0.011 $mm$ respectively. With this distribution one can expect to obtain approximately 1000 grains for every 1 $mm^2$ of surface. As such, the simulation with 50000 grains fits in a domain with 50 $mm^2$ and the one with 560000 grains in a domain of 560 $mm^2$ of surface. After the generation of the Laguerre-Voronoi tessellation, the real number of grains were 52983 and 544913 grains for the domains of 50 and 560 $mm^2$ respectively. \\

\subsubsection{ 2D grain growth 50000 Initial grains}

\begin{figure}[!h]
\centering
\includegraphics[width=1.0\textwidth] {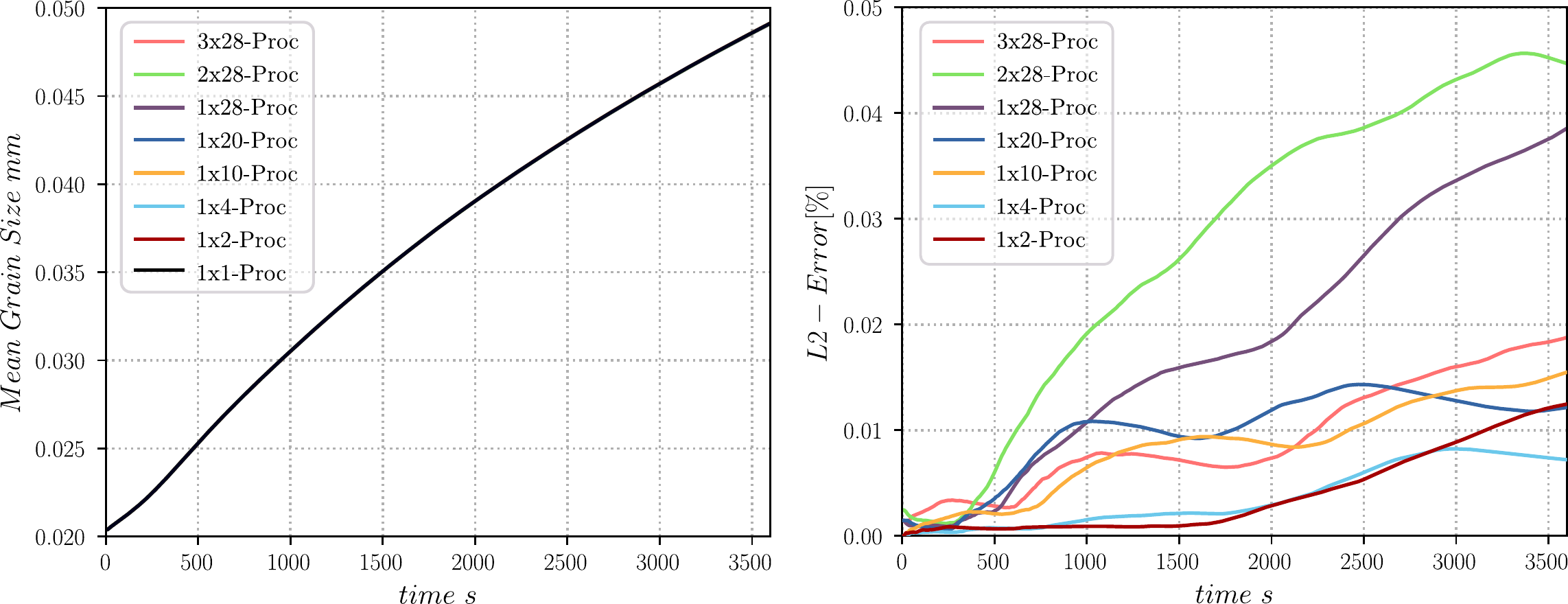}
\caption{ Results of the test with a surface of 50 $mm^2$ performed in 1, 2, 3 and 4, 10, 20, 28, 56 (2x28), and 84 (2x28) processors. Here the mesh size parameter is the same for all runs and equivalent to $h_{trm}=0.004$ $mm$ and the time step is $dt=10$ $s$. Left: Mean grain size evolution, right: L-2 Error of the evolution of the Mean Size with the test performed in sequential (1 processor) as a reference.}
\label{fig:MeanSize_50000}
\end{figure}

Multiple simulations for the case with 50 $mm^2$ of surface were performed on 1 (sequential), 2, 4, 10, 20, 28, 56 (2x28) and 84 (3x28) processors, the mesh size parameter and the time step will be held constant and equal to to $h_{trm}=0.004$ $mm$ and $dt=10$ $s$ respectively.\\

\begin{figure}[!h]
\centering
\includegraphics[width=0.8\textwidth] {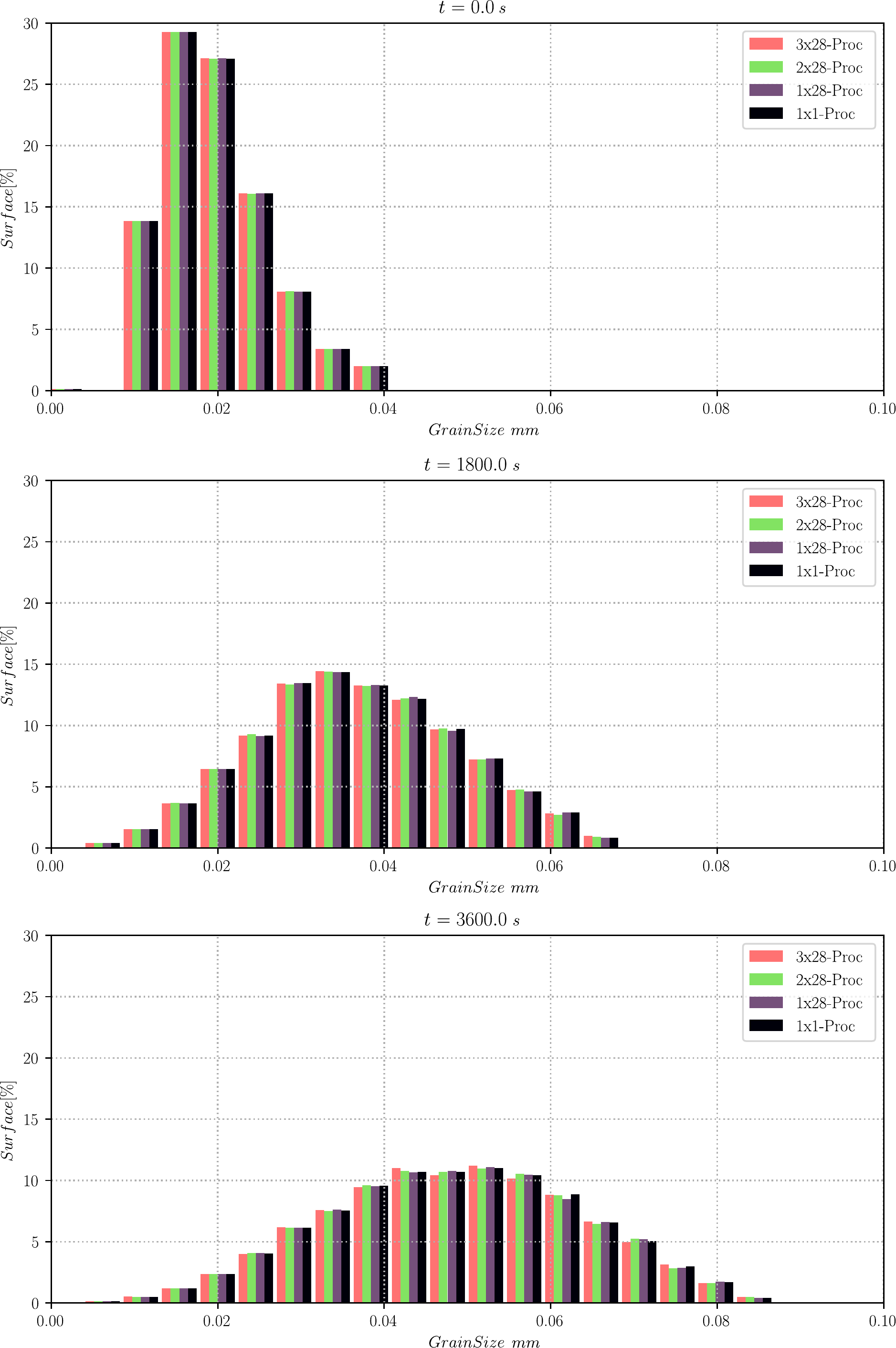}
\caption{ Results of grain size distribution for the test with a surface of 50 $mm^2$ performed in 1, 28, 56 (2x28) and 84 (3x28) processors. Initial state (top), distributions after 1800 $s$ (center), distributions after 3600 $s$ (bottom).}
\label{fig:Histograms_50000}
\end{figure}

\begin{figure}[!h]
\centering
\includegraphics[width=0.7\textwidth] {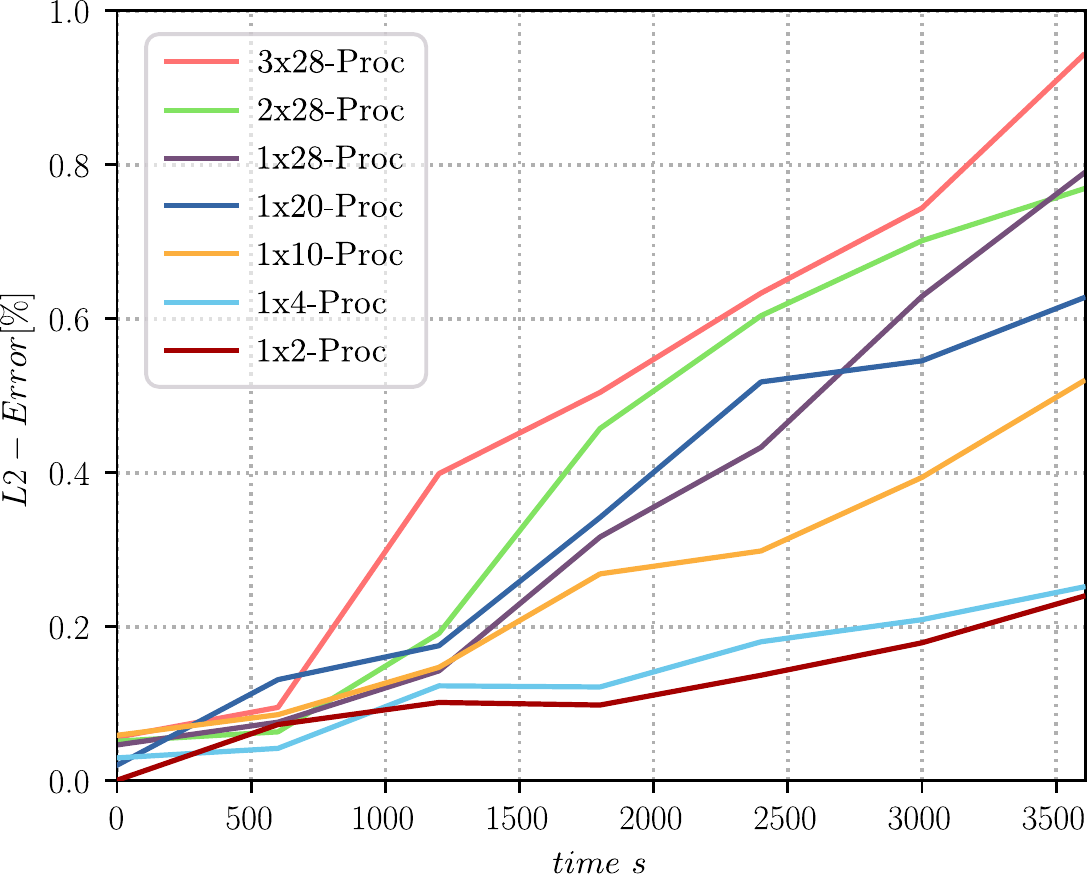}
\caption{ L2-Error over the grain size distribution for the test with a surface of 50 $mm^2$ performed in 2, 4, 10, 20, 28, 56 (2x28) and 84 (3x28) processors compared to the simulation performed in 1 processor.}
\label{fig:Hist_Error_50000}
\end{figure}

\begin{figure}[!h]
\centering
\includegraphics[width=1.0\textwidth] {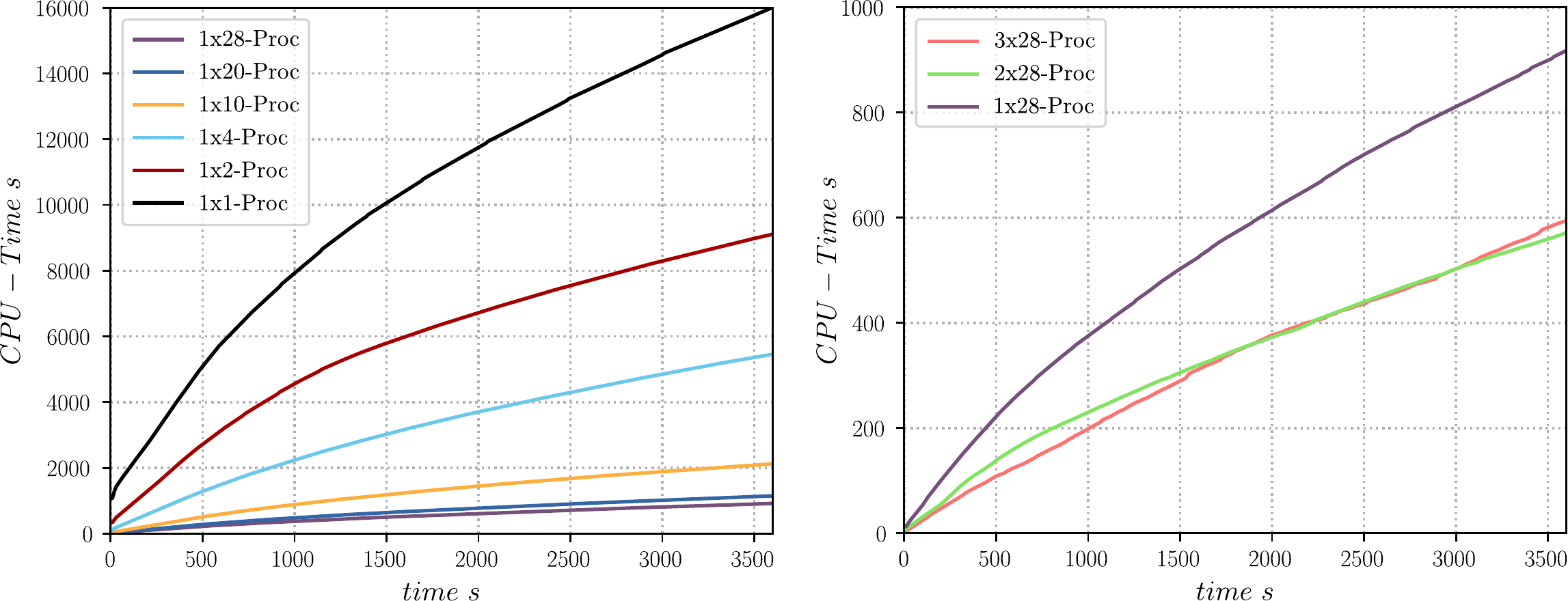}
\caption{ CPU-times for the test with a surface of 50 $mm^2$ performed in 2, 4, 10, 20 and 28 processors (left) and 28, 56 (2x28) and 84 (3x28) processors (right).}
\label{fig:CPU-Time_50000}
\end{figure}

Figure \ref{fig:MeanSize_50000} gives the results for the evolution of mean grain size pondered by surface and the L2-Error (mesured with respect to the sequential case) for the test with 50 $mm^2$ of surface. Here, all parallel tests behave almost exactly as the sequential one (1 processor). The L2-Error is found to be lower than 0.05 \% in relation to the test performed in sequential. Similarly, figure \ref{fig:Histograms_50000} shows the evolution of the mean grain size distributions performed in 1, 28, 56 and 84 processors. Once again the results are almost identical for all tests. This is illustrated in Figure \ref{fig:Hist_Error_50000} which shows that the error over the grain size distributions in surface is in all cases lower than 1.0 \%. Furthermore, Figure \ref{fig:CPU-Time_50000} shows the CPU-time needed for all the configurations to accomplish the test. Here, the parallel tests obtained a speed-up\footnote{The speed up in the \emph{Variable Charge} test case (strong scaling benchmark) is obtained by dividing the CPU-time of the sequential case by the CPU-time of the parallel case in question.}, being the test with 56 processor the fastest one, contrary to the expected behaviour where the fastest test should be the one performed with 84 processors. In fact, the simulation performed over 84 processors may be over-partitionned, meaning that the number of partitions is too high compared to the number of elements in the simulation.\\

\begin{figure}[!h]
\centering
\includegraphics[width=1.0\textwidth] {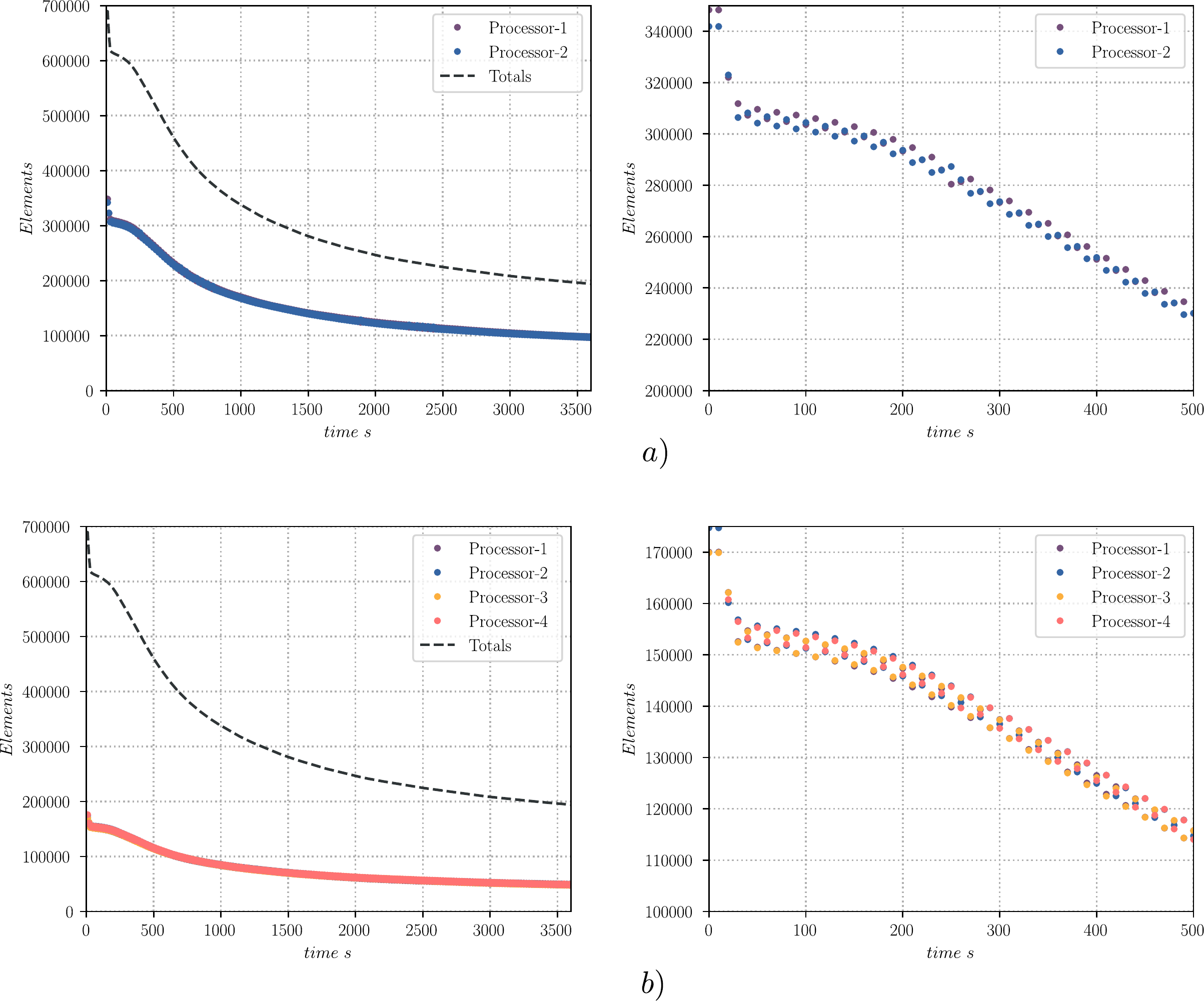}
\caption{Number of elements per processor for the for the test with a surface of 50 $mm^2$ performed in 2 and 4 processors compared to the total number. The evolution of the number of elements (left) and a zoom at the begining of the simulation is shown (right). a) 2 Processors, b) 4 processors.}
\label{fig:Elements_FewProcs}
\end{figure}

\begin{figure}[!h]
\centering
\includegraphics[width=1.0\textwidth] {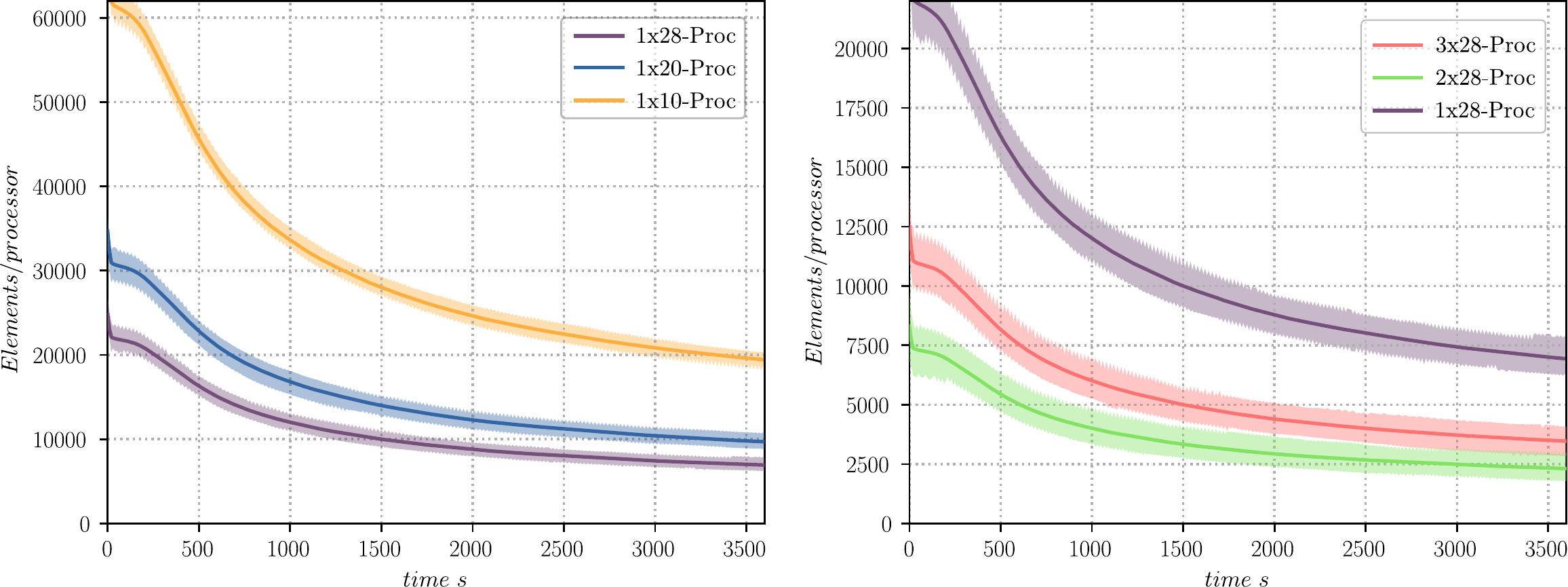}
\caption{Mean number of elements per processor for the test with a surface of 50 $mm^2$ performed in 10, 20, 28, 56 (2x28) and 84 (3x28) processors. The evolution of the number of elements for the simulations contained in one node (left) and in multiple nodes (right) is shown. The range for the number of elements of all processors in the same simulation is shown in the same color with an alpha component.}
\label{fig:EltNumbers_50000}
\end{figure}

Figure  \ref{fig:Elements_FewProcs} illustrates 2 examples of the evolution of elements on each partition for 2 cases: 2 processors and 4 processors. Note that in this case (the parallel case, contrary to the sequential one), the response is not a curve but a range of data in the Y axis, as the amount of Elements on each partitions is changing every time the mesh scattering is performed. Note that this range is higher when the number of processors increases, as the boundaries between partitions present more changes, however for this two examples this range is well contained around $1/2$ and $1/4$ of the total number of elements, for the simulation performed with 2 and 4 processors respectively. The re-equilibrium of charges here is optimal. Figure \ref{fig:EltNumbers_50000} gives the mean number of elements and its range for the simulations performed in 10, 20, 28, 56 and 84 processors, here two plots are given: for the simulations performed in one CPU (one node) and in multiple CPUs (multiple nodes).  \\

\begin{figure}[!h]
\centering
\includegraphics[width=1.0\textwidth] {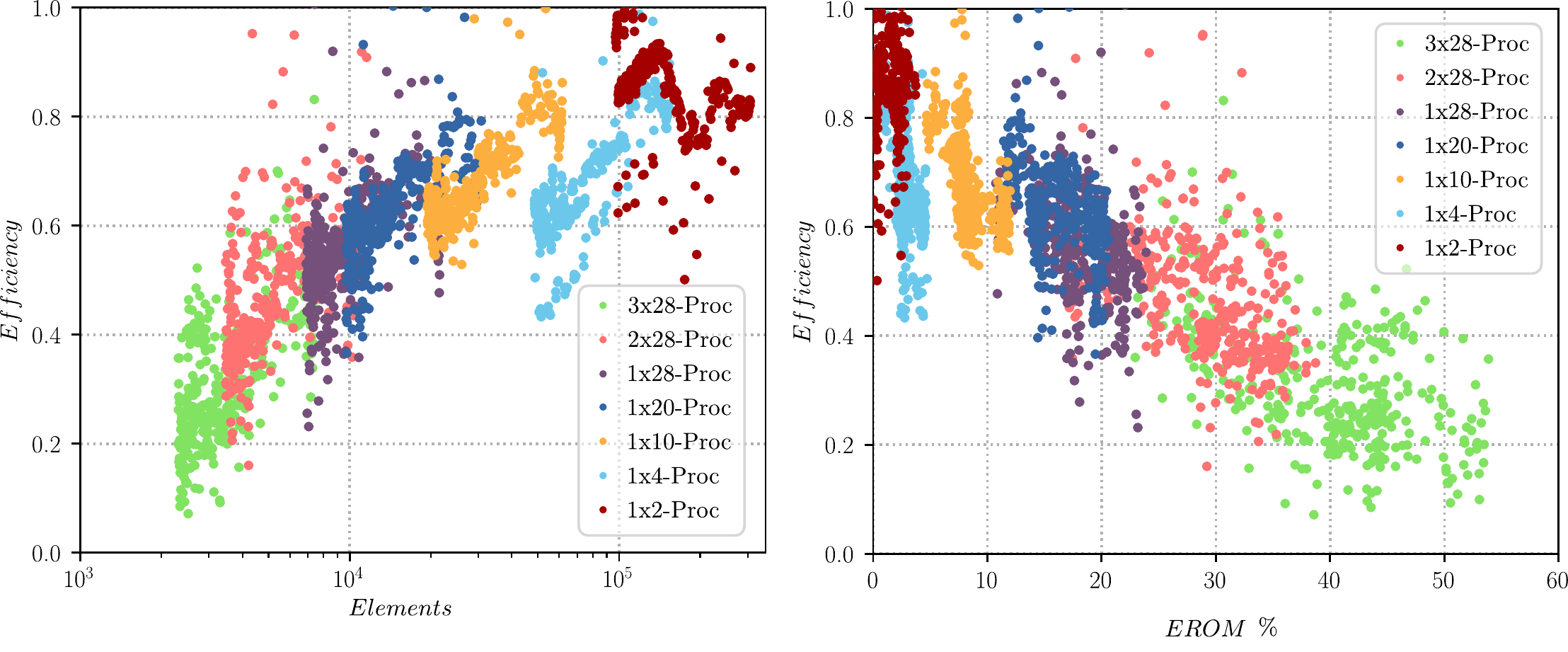}
\caption{Efficiency of the simulation for the case with a surface of 50 $mm^2$ against the mean number of elements (left) and against the $EROM$ index (right).}
\label{fig:Eff_EltNumbers}
\end{figure}

One interesting index to follow is the value obtained by dividing the range of elements over its mean value, hereafter called the index $EROM$. In parallel computations, this index have been plotted in figure \ref{fig:RangePercentPlot} and shows how much the elements may be scattered relatively to the mean number of elements on each partition. The index $EROM$ reflects the maximum limit percent of elements being transported from one partition to another by the Mesh Scattering procedure presented in section \ref{sec:repartitionning}, hence one can use it to study the global efficiency of the numerical procedure. Here we define the efficiency of one increment $i$ for a simulation with $N_{p}$ number of processors as follows: 

\begin{equation}
\label{Eq:Efficiency_Increment}
\centering
Efficiency=\dfrac{t^i_{1}} { t^i_{N_{p}} \cdot N_{p}} 
\end{equation}

where the term  $t^i_{N_{p}}$ determines the CPU-time needed to perform increment $i$ in a simulation with $N_{p}$ number of processors. This equation computes the amount of resources needed for an increment of a parallel simulation compared to a sequential one. Figure \ref{fig:Eff_EltNumbers} plots the Efficiency of the simulation for the case with 50000 grains against the mean number of elements (left) and against our $EROM$ index (right). Of course the lower the efficiency per increment for a given simulation, the lower we expect to be the speed-up of the parallel simulations. The mean efficiency of the 84 processors test (approximately 0.29) is much lower than the mean efficiency of the 56 processors test (approximately 0.48), thus a reduction of the efficiency of 65\% for an increase in the number of processors of 50\% (from 56 to 84 processors). These results suggest that the simulation performed with 84 processors is indeed over partitionned and that one should aim to obtain a mean number of elements not lower than 10000 or a $EROM$ index higher than 30\% (The real impact of the $EROM$ index will be studied in the case at constant charge). Note however, that the relative overhead obtained by the high number of partitions is not higher than the reduction in CPU-time obtained, as the speed-up is still greater than 1 even for the simulation with 84 processors.\\

\begin{figure}[!h]
\centering
\includegraphics[width=0.7\textwidth] {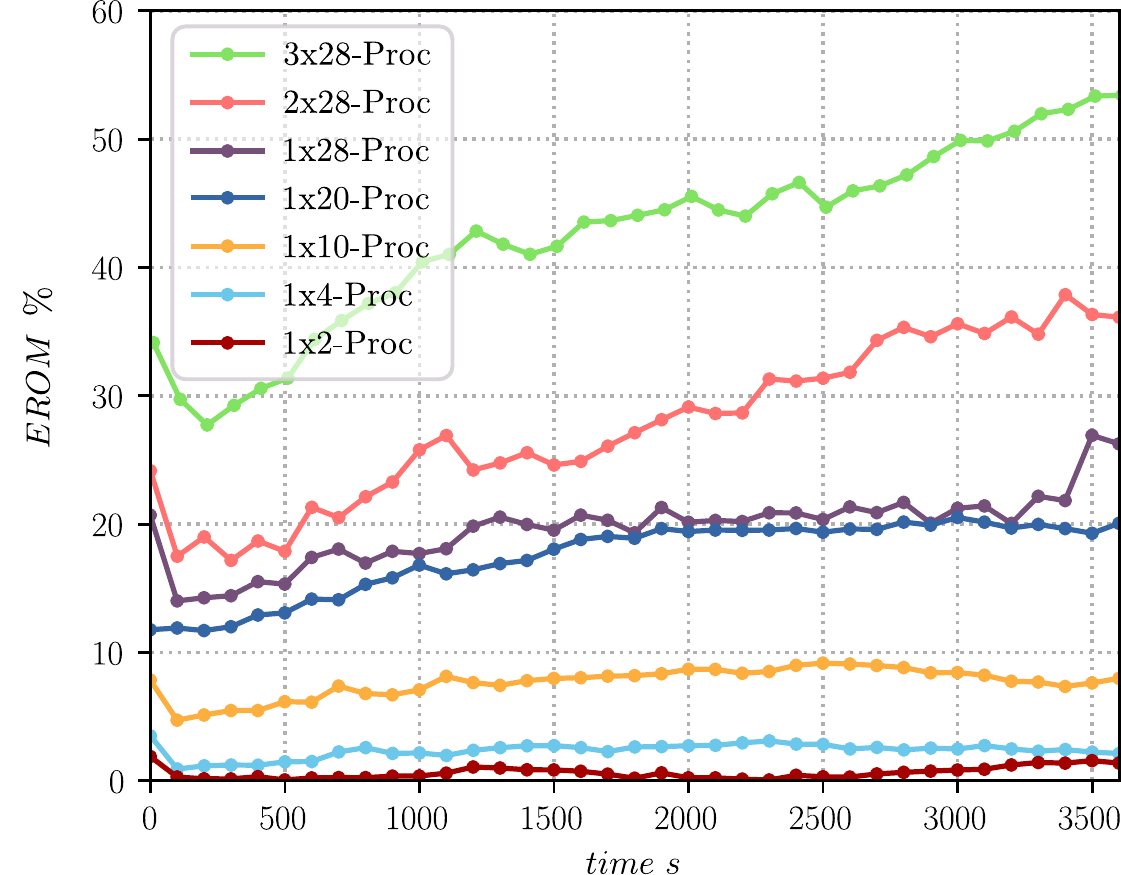}
\caption{Range of the number of elements over the Mean number of elements per processor (index $EROM$) for the test with a surface of 50 $mm^2$ performed in 10, 20, 28, 56 (2x28) and 84 (3x28) processors. The percent value increases with the number of processors plotted every 10 increments.}
\label{fig:RangePercentPlot}
\end{figure}

Finally Figure \ref{fig:SpeedUp_CpuTime} plots the speed-up of the parallel implementation of the TRM model against the number of processors, here the reference for the optimal speed-up is also shown. Of course the optimal speed-up can not be obtained in our case as the operations produced by the communication create additional overhead.

\subsubsection{ 2D grain growth 560000 Initial grains}

\begin{figure}[!h]
\centering
\includegraphics[width=1.0\textwidth] {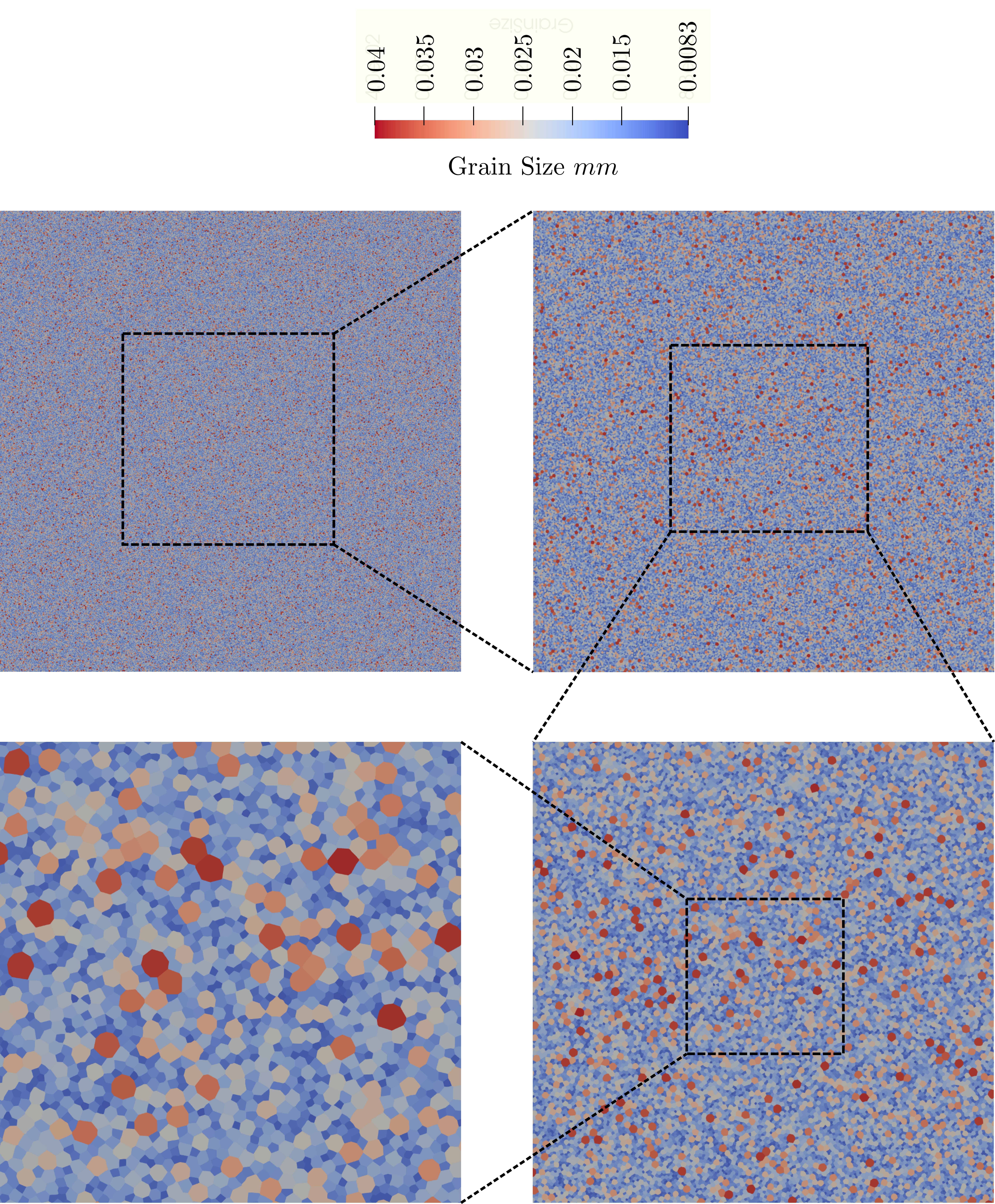}
\caption{ Initial state of the case with a surface of 560 $mm^2$, 544913 grains are shown in the bigger image. Subsequent zoom views are given.}
\label{fig:560000GrainsShots}
\end{figure}

Here a similar GG test will be performed but the surface of the domain will be increased to $560 mm^2$. The test will be performed on 1, 14, 28, 56 (2x28), 84 (3x28), 112 (4x28) and 140 (5x28) processors. Figure \ref{fig:560000GrainsShots} illustrates the initial microstructure for this case. To the knowledge of the author, this is the maximum amount of grains that have been attempted to simulate using 2D unstructured Finite-Element meshas, and the second largest simulation in the context of GG, the first is the one presented in \cite{Elsey2013} for a microstructure with 671000 initial grains, but in a Fast Fourier Transform (FFT) context (Thus using regular grids). \\

\begin{figure}[!h]
\centering
\includegraphics[width=1.0\textwidth] {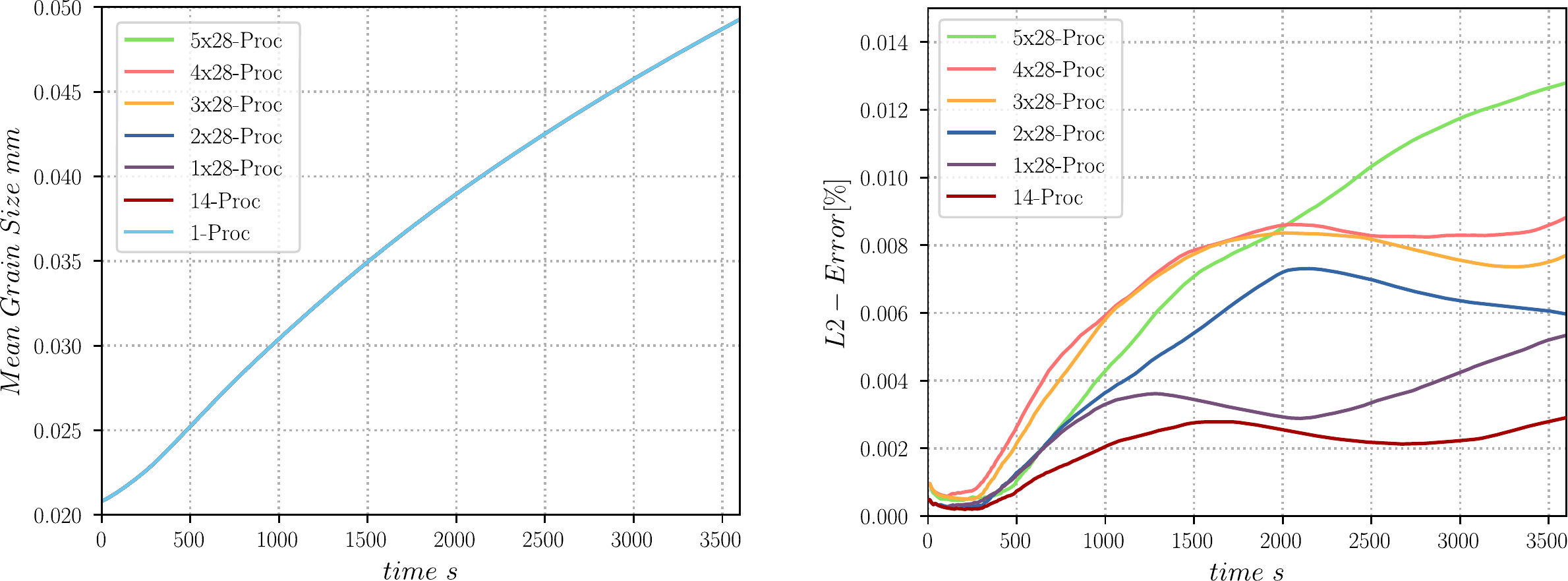}
\caption{ Results of the case with a surface of 560 $mm^2$ performed in 1, 14, 28, 56 (2x28), 84 (3x28), 112 (4x28) and 140 (5x28) processors. Here the mesh size parameter is the same for all runs and equivalent to $h_{trm}=0.004$ $mm$ and the time step is $dt=10$ $s$. left: Mean grain size evolution, right: L-2 Error of the evolution of the Mean Size with the test performed in sequential (1 processor) as a reference.}
\label{fig:MeanSize_560000}
\end{figure}

\begin{figure}[!h]
\centering
\includegraphics[width=0.7\textwidth] {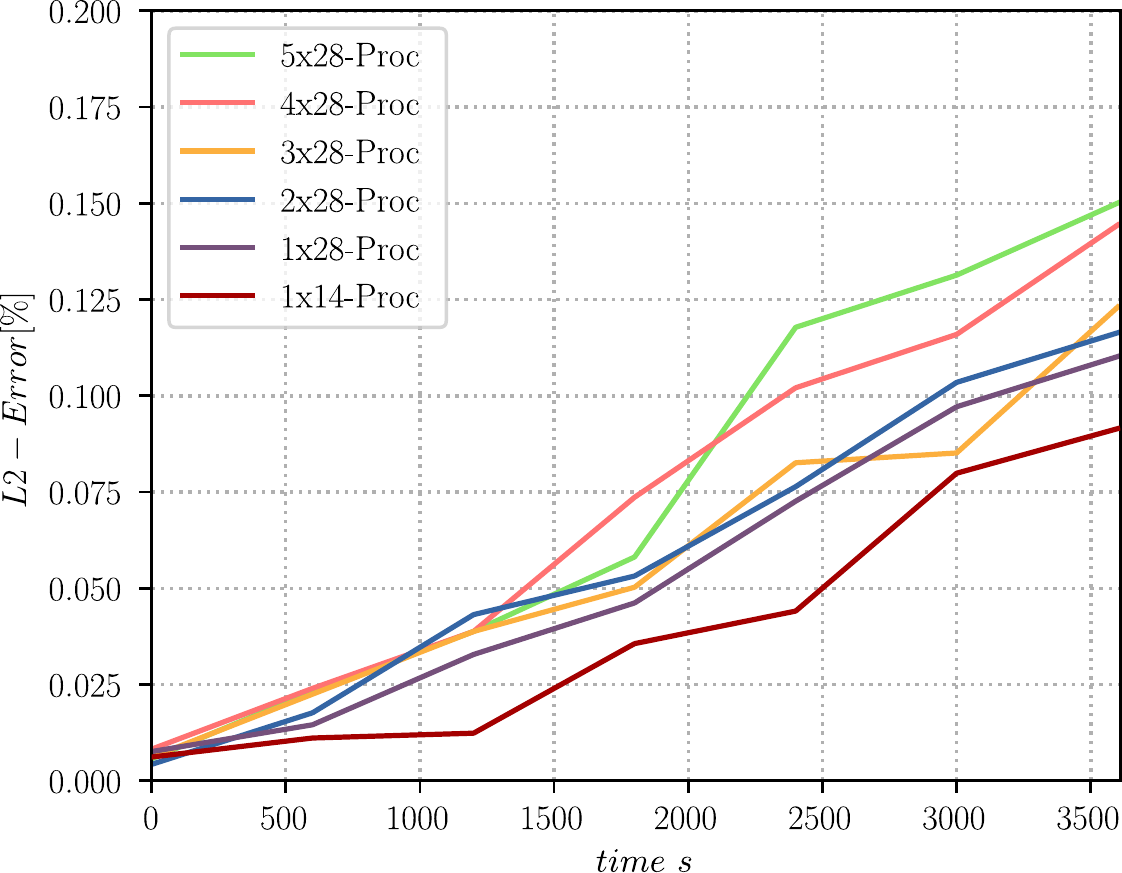}
\caption{ L2-Error over the grain size distribution for the case with a surface of 560 $mm^2$ performed in 14, 28, 56 (2x28), 84 (3x28), 112 (4x28) and 140 (5x28) processors compared to the simulation performed in 1 processor.}
\label{fig:Hist_Error_560000}
\end{figure}

Figure \ref{fig:MeanSize_560000} gives the results for the evolution of mean grain size and its L2-error for the test with 560 $mm^2$ of surface. Similarly, figure \ref{fig:Hist_Error_560000} plots the L2-Error over the grain size distribution of the present test. The range of both L2-Errors have been reduced by at least 1/5 compared to the tests with 50 $mm^2$ of surface. At this level we consider that the influence of the parallel TRM model over the precision of the simulation is quasi-non-existent, while for the test with 50 $mm^2$ of surface this influence is very low.\\

\begin{figure}[!h]
\centering
\includegraphics[width=0.7\textwidth] {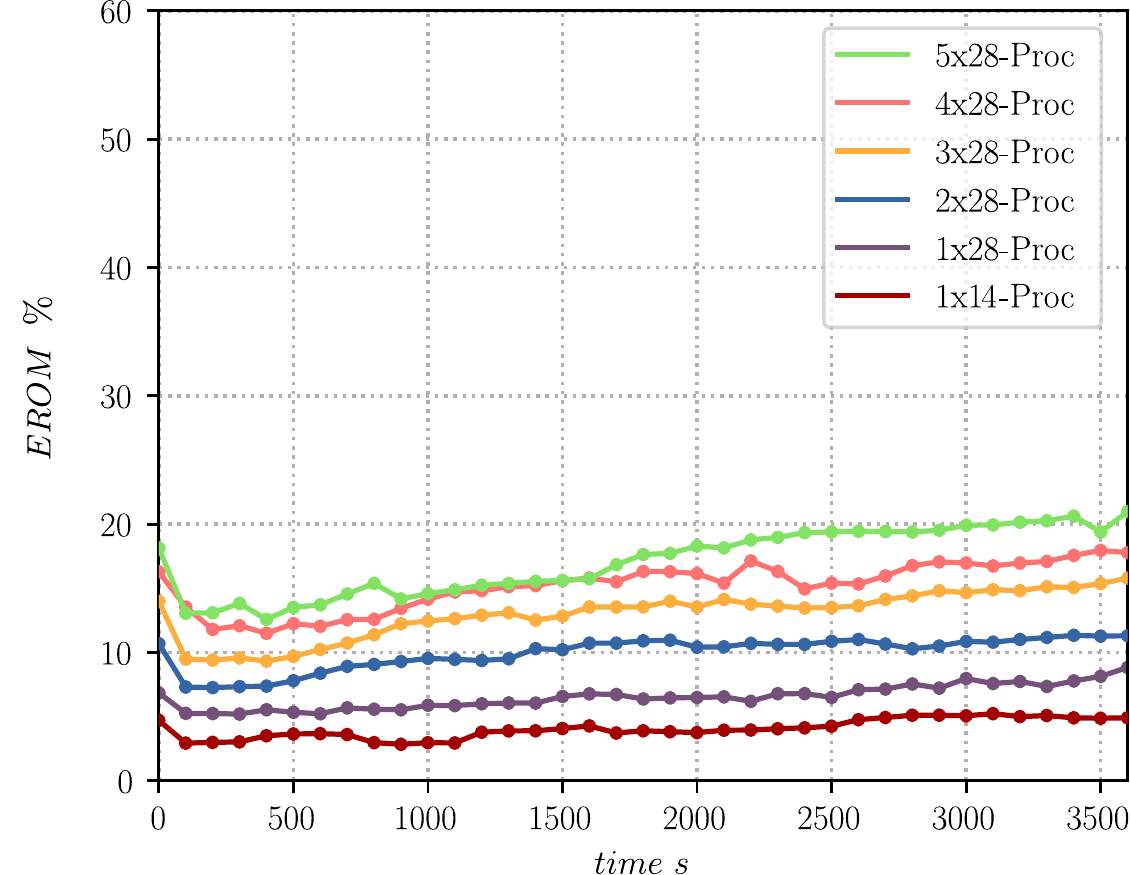}
\caption{Range of the number of elements over the Mean number of elements per processor (index $EROM$) for the test with 560 $mm^2$ of surface performed in 14, 28, 56 (2x28), 84 (3x28), 112 (4x28) and 140 (5x28) processors. The percent value increases with the number of processors plotted every 10 increments.}
\label{fig:RangePercentPlot}
\end{figure}

By increasing the simulated domain size, the index $EROM$ should be reduced as the mean number of elements per processor is increased. Figure \ref{fig:RangePercentPlot} shows the evolution of the index $EROM$ for the present test, here the evolution of this index has been reduced by more than half (considering the simulation with 3x28 in both the 50000 grains and the 560000 grains case) which maintains the efficiency of a time step over 0.4 accordingly to figure \ref{fig:Eff_EltNumbers}.\\

\begin{figure}[!h]
\centering
\includegraphics[width=0.8\textwidth] {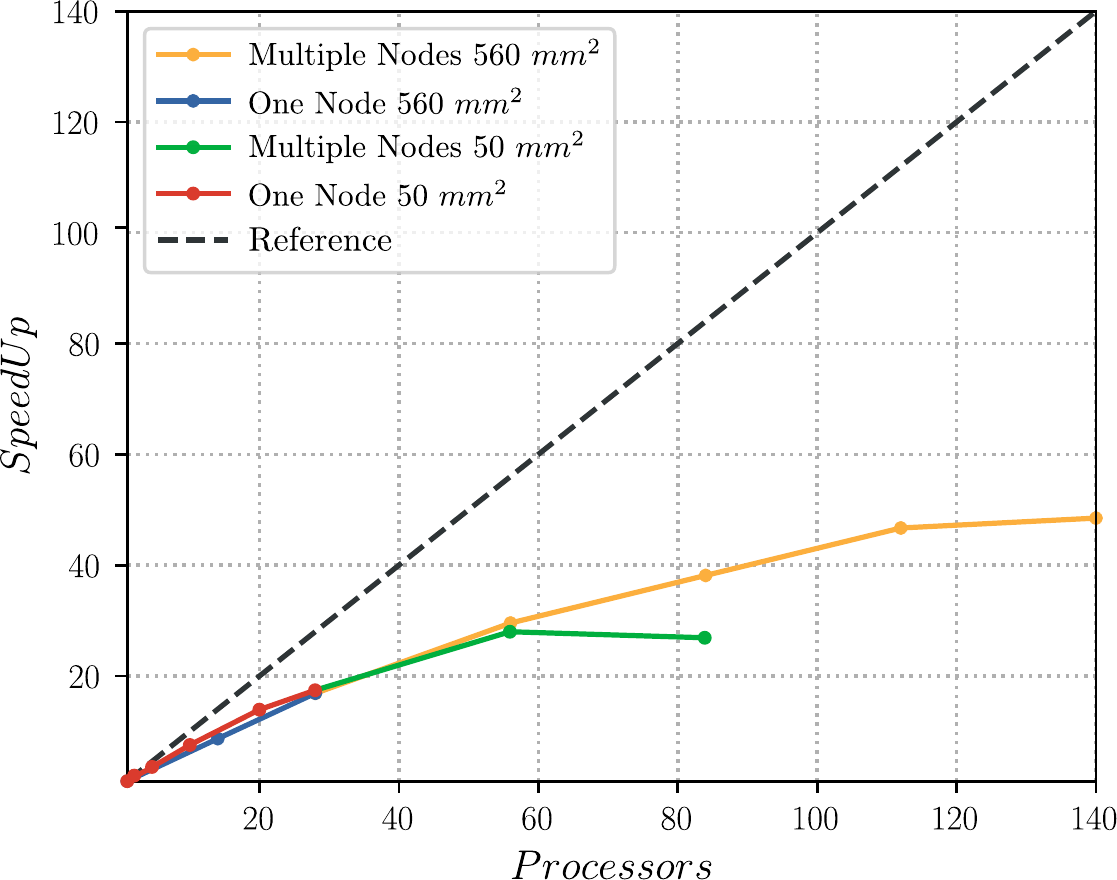}
\caption{Performance of the TRM model in parallel for a variable processor charge: the domain surface is maintained constant while the number of processors increases, two test were performed, one with a surface of 50 $mm^2$ and the second with a surface of 560 $mm^2$. The data is compared to the optimal speed up (Reference).}
\label{fig:SpeedUp_CpuTime}
\end{figure}

Figure \ref{fig:SpeedUp_CpuTime} illustrates the evolution of the speed up of both test cases. This result shows that the speed up is relatively the same up to 56 (2x28) processors but it diverge for a higher number of processors in favor of the simulation with 560 $mm^2$ of surface. The maximum speed-up obtained was for the case with a surface of 560 $mm^2$ performed over 140 (5x28) processors for which a speed-up of 48.5 was obtained, this simulation took 38 minutes and 45 seconds while the one performed in 1 processor took 31 hours and 20 minutes. The total number of grains at the end of this simulation was of 117157 grains, hence 21.5 $\%$ of the total initial number of grains while for the simulation with 50 $mm^2$ of surface was of 10399, hence 19.6 $\%$ of the total initial number of grains.

\subsection{Constant Processor Charge (weak scaling benchmark)}\label{ConstantTest}

\begin{figure}[!h]
\centering
\includegraphics[width=1.0\textwidth] {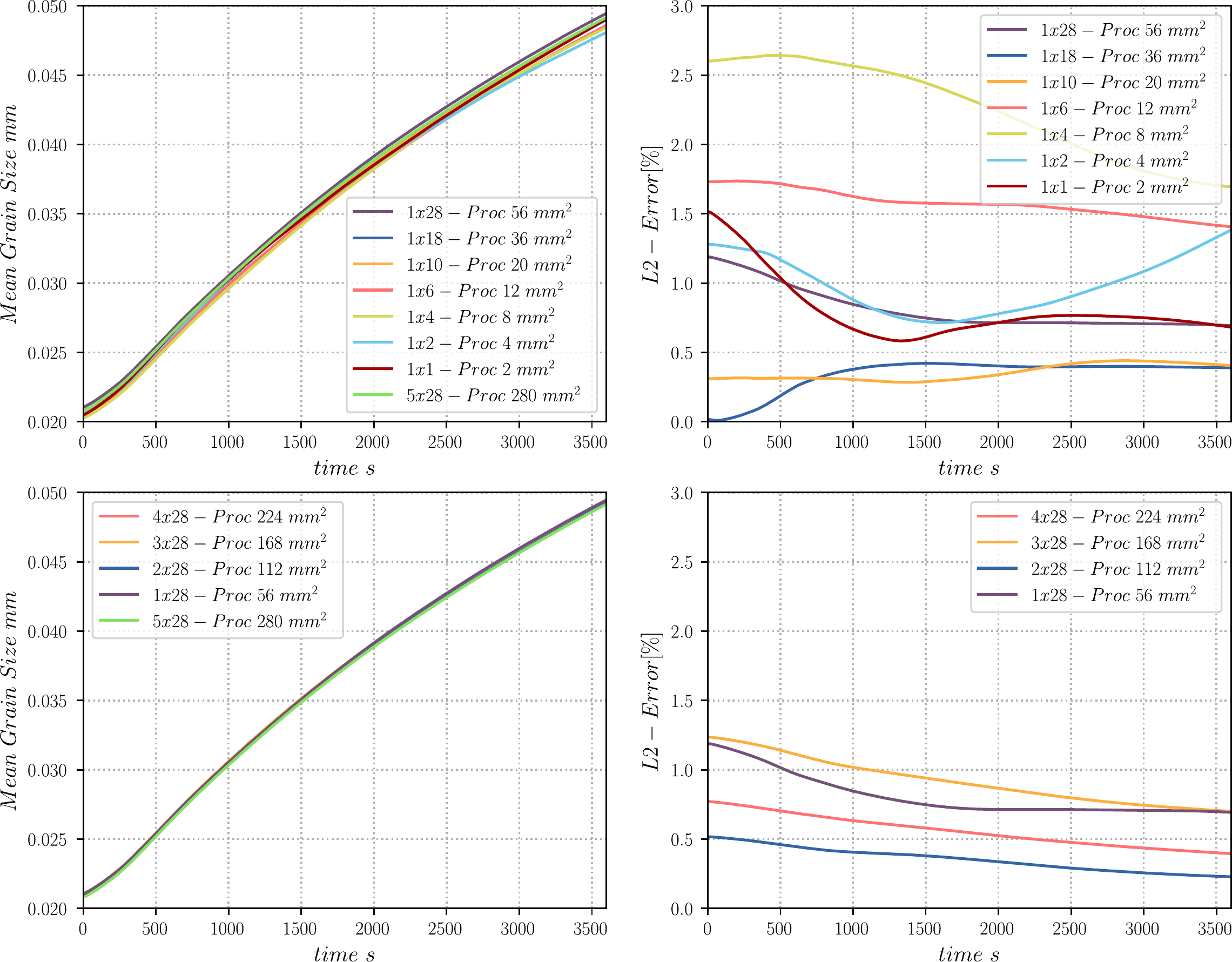}
\caption{ Mean size results of the case with a DSPP of 2 $mm^2$ performed in 1, 2, 4, 6, 10, 18, 28, 56 (2x28), 84 (3x28), 112 (4x28) and 140 (5x28) processors. The plot has been divided in two plots for a clearer visibility: top: simulations performed over one Node (1, 2, 4, 6, 10, 18, 28), bottom: simulations performed over multiple Nodes (28, 56 (2x28), 84 (3x28), 112 (4x28), 140 (5x28)). Each plot is given its respective L2-Error to the response of the largest simulation in this context (140 (5x28) processors).}
\label{fig:MeanSize_Progressive_2000}
\end{figure}

When performing a benchmark on a strong scaling context, the amount of memory that a processor has to maintain decreases when the amount of processors increase. Of course in the same context, when performing a sequential simulation all the memory has to be maintained by only one processor, making it longer to read or to add informations to the data set of the running process. This artifact makes it, that the optimization made by the parallel implementation in a strong scaling context, be both, for memory managment (to acces and to write in a given memory location) and in number of operations (such as remeshing or moving nodes) to perform by a processor. On the other hand, the aim of a benchmark at constant processor charge is to measure the speed-up\footnote{contrary to the speed up in the \emph{Variable Charge} test case (strong scaling benchmark), the speed up for the \emph{Constant Charge} test case is obtained by multiplying the CPU-time of the sequential case with the number of processors of the parallel case, divided by the CPU-time of the parallel case.} generated by a parallel implementation only on the number of operations to be performed inside a processor, as the memory to be maintained by every processor is constant when the number of processors increases. While it is not possible to maintain an equally distributed and constant charge on all processors here, it is possible to approach the concept by increasing the size of the simulated domain proportionally to the number of processors used in a parallel simulation with the TRM model. Two sets of simulations will be performed: the first with a domain surface per processor (DSPP) of 2 $mm^2$ (approximately 2000 grains per processor) and the second with a DSPP of 4 $mm^2$ (approximately 4000 grains per processor). Moreover, even though the same grain size distribution is used in the generation of the Laguerre-Voronoi tessellation, it is not possible to obtain a perfectly equal statistical distribution for different sizes of domains. As such, at the beginning of the simulations small variations on the mean grain size or the grain size distributions may appear.\\

\begin{figure}[!h]
\centering
\includegraphics[width=1.0\textwidth] {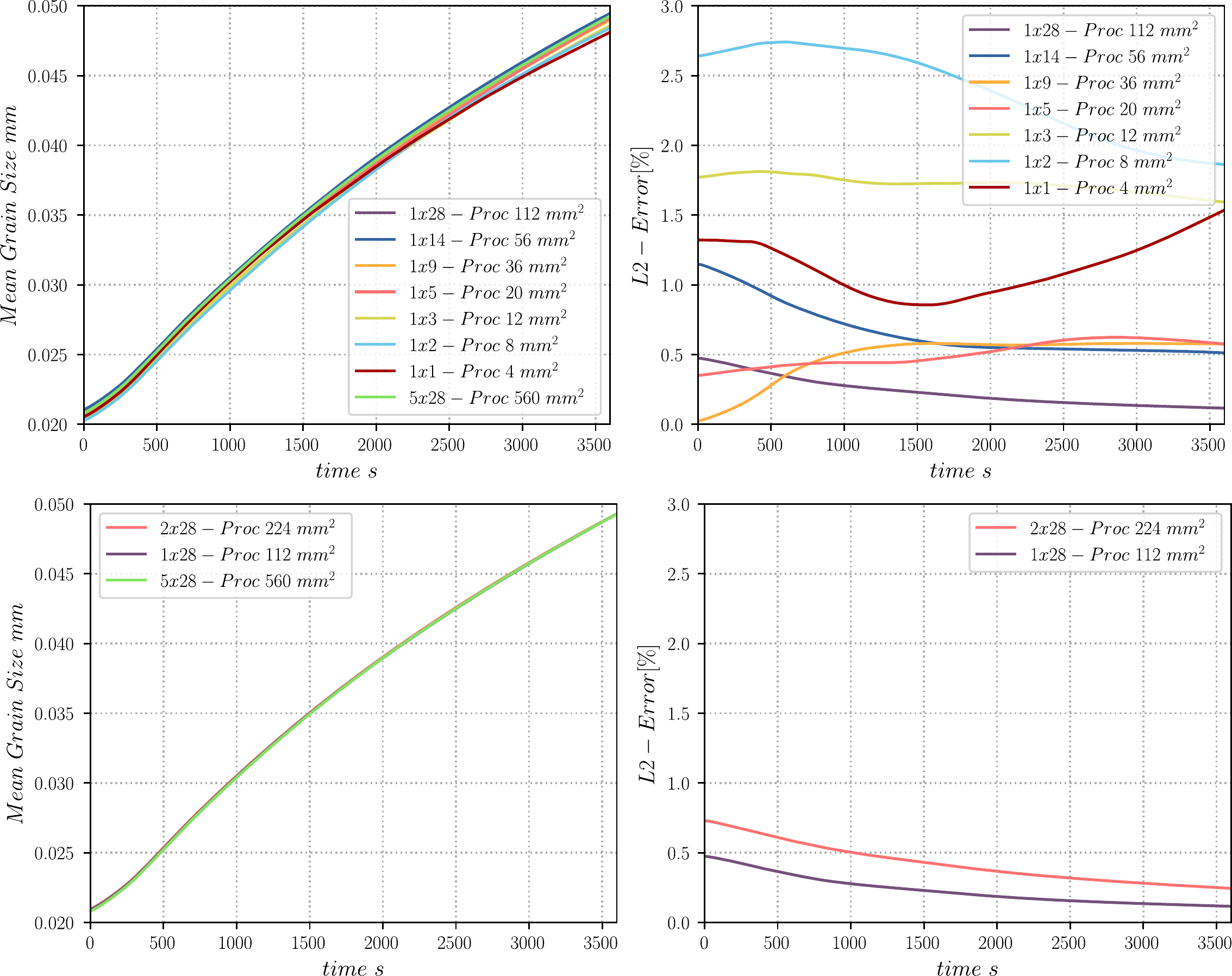}
\caption{ Mean size results of the case with a DSPP of 4 $mm^2$ performed in 1, 2, 3, 5, 9, 14, 28, 56 (2x28) and 140 (5x28) processors. The plot has been divided in two plots for a clearer visibility: top: simulations performed over one Node (1, 2, 4, 6, 10, 18, 28), bottom: simulations performed over multiple Nodes (28, 56 (2x28), 140 (5x28)). Each plot is given its respective L2-Error to the response of the largest simulation in this context (140 (5x28) processors).}
\label{fig:MeanSize_Progressive_4000}
\end{figure}

Figure \ref{fig:MeanSize_Progressive_2000} and \ref{fig:MeanSize_Progressive_4000} give the results for the evolution of the mean size of the cases with a DSPP of 2 $mm^2$ and 4 $mm^2$ respectively. Furthermore, L2-Error plots are also given corresponding to the difference of the simulations performed in parallel to the reference cases (here chosen as the largest simulation of each context which were performed in 140 (5x28) processors: 280 and 560 $mm^2$ respectively for the cases with a DSPP of 2 $mm^2$ and 4 $mm^2$). In both cases the curves appear to be very similar to their references with an error inferior to 3\%. Note that the L2-Error have a tendency to decrease when the simulation domain increases (as expected), and that the error after a total surface of 56 $mm^2$ (approximately 56000 initial grains) is inferior to 1\%.\\


\begin{figure}[!h]
\centering
\includegraphics[width=0.8\textwidth] {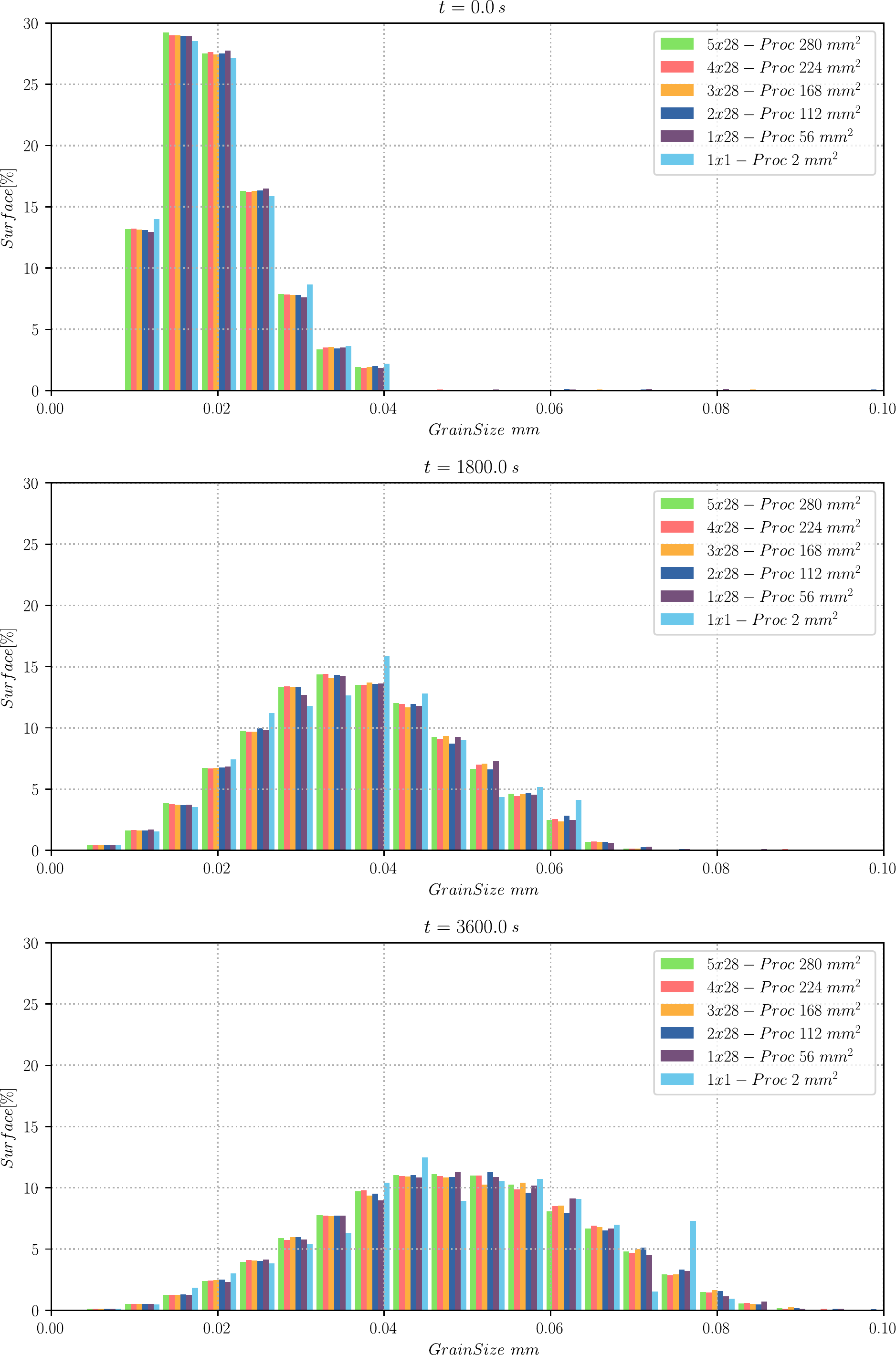}
\caption{ Results of grain size distribution for the test with a DSPP of 2 $mm^2$ performed in 1, 28, 56 (2x28), 84 (3x28), 112 (4x28) and 140 (5x28) processors. Initial state (top), distributions after 1800 $s$ (center), distributions after 3600 $s$ (bottom).}
\label{fig:Histo_Progressive_2000}
\end{figure}

\begin{figure}[!h]
\centering
\includegraphics[width=1.0\textwidth] {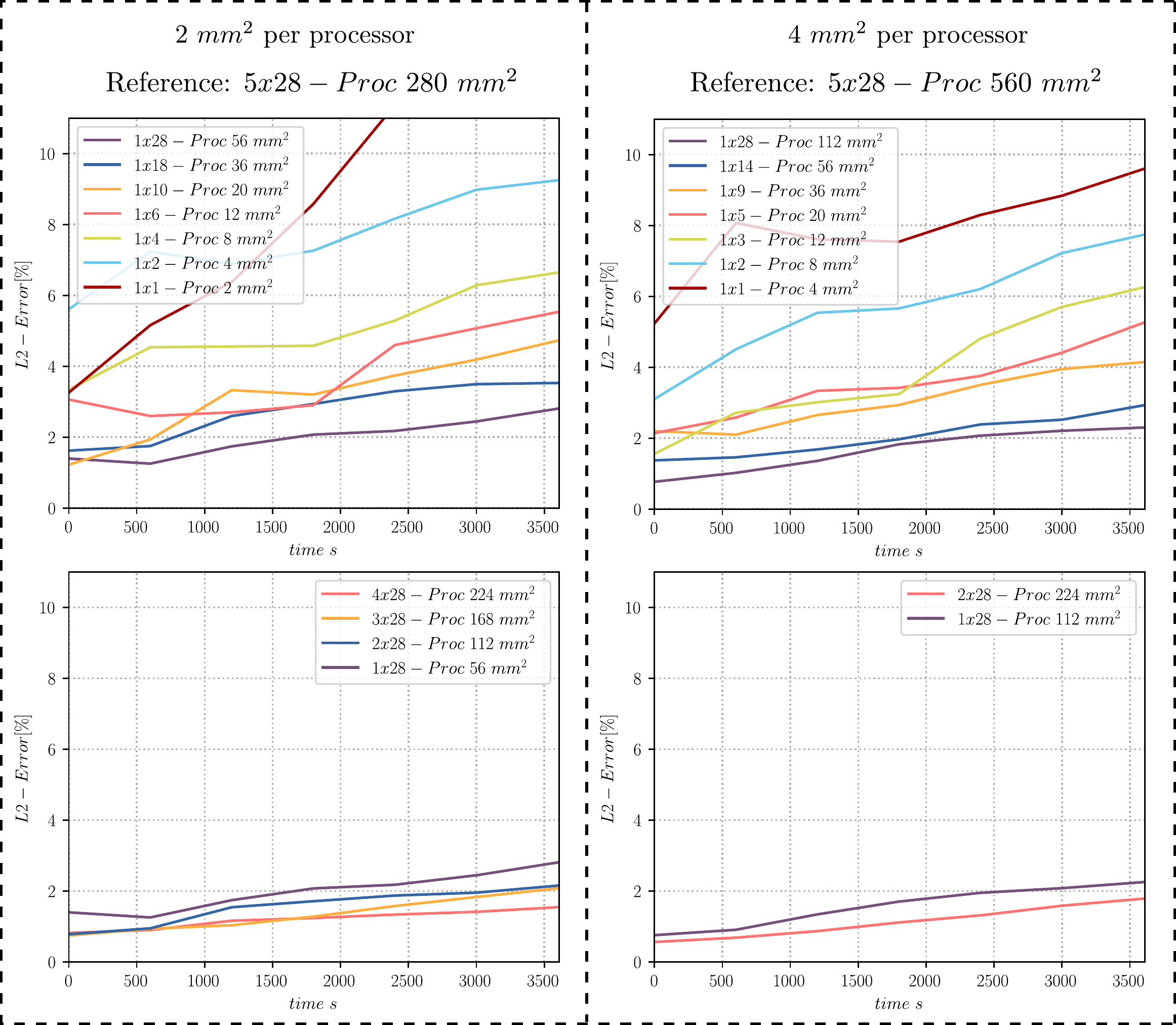}
\caption{L2-Error of the grain size distribution for the test with a DSPP of 2 $mm^2$ (left) and a DSPP of 4 $mm^2$ (right) compared to of the largest simulation of each context (140 (5x28) processors) (280 and 560 $mm^2$ respectively). Each plot have been divided in two for a clearer visibility: top: simulations performed over one Node, bottom: simulations performed over multiple Nodes.}
\label{fig:Histo_Error_Progressive}
\end{figure}

Figure  \ref{fig:Histo_Progressive_2000} illustrates the evolution of the mean size distribution in surface for one set of simulations with a DSPP of 2 $mm^2$, at the begining, after 1800 $s$ and after 3600 $s$ of simulated time, here only the simulations performed in 1, 28, 56, 84, 112, and 140 processors are plotted for a clearer visualization. Figure \ref{fig:Histo_Error_Progressive} gives the evolution of the L2-Error over the grain size distributions for both sets of simulations with a DSPP of 2 $mm^2$ and of 4 $mm^2$ and for all the simulations performed. Similarly to the evolution of the mean grain size, the L2-Error is obtained to be lower than 3\% after a domain size of 56 $mm^2$, suggesting that our microstructure can be statistically well represented by a simulation with over 56000 initial grains. On the contrary, simulations performed with a domain of 20 $mm^2$ and below, show an error at the end of the simulation of more than 5\% which suggest that they contain too few grains (at the end of the simulation) or that statistical results may be highly influenced by the boundary conditions, simulations with lower than 20000 grains should be avoided in our GG context.\\


\begin{figure}[!h]
\centering
\includegraphics[width=1.0\textwidth] {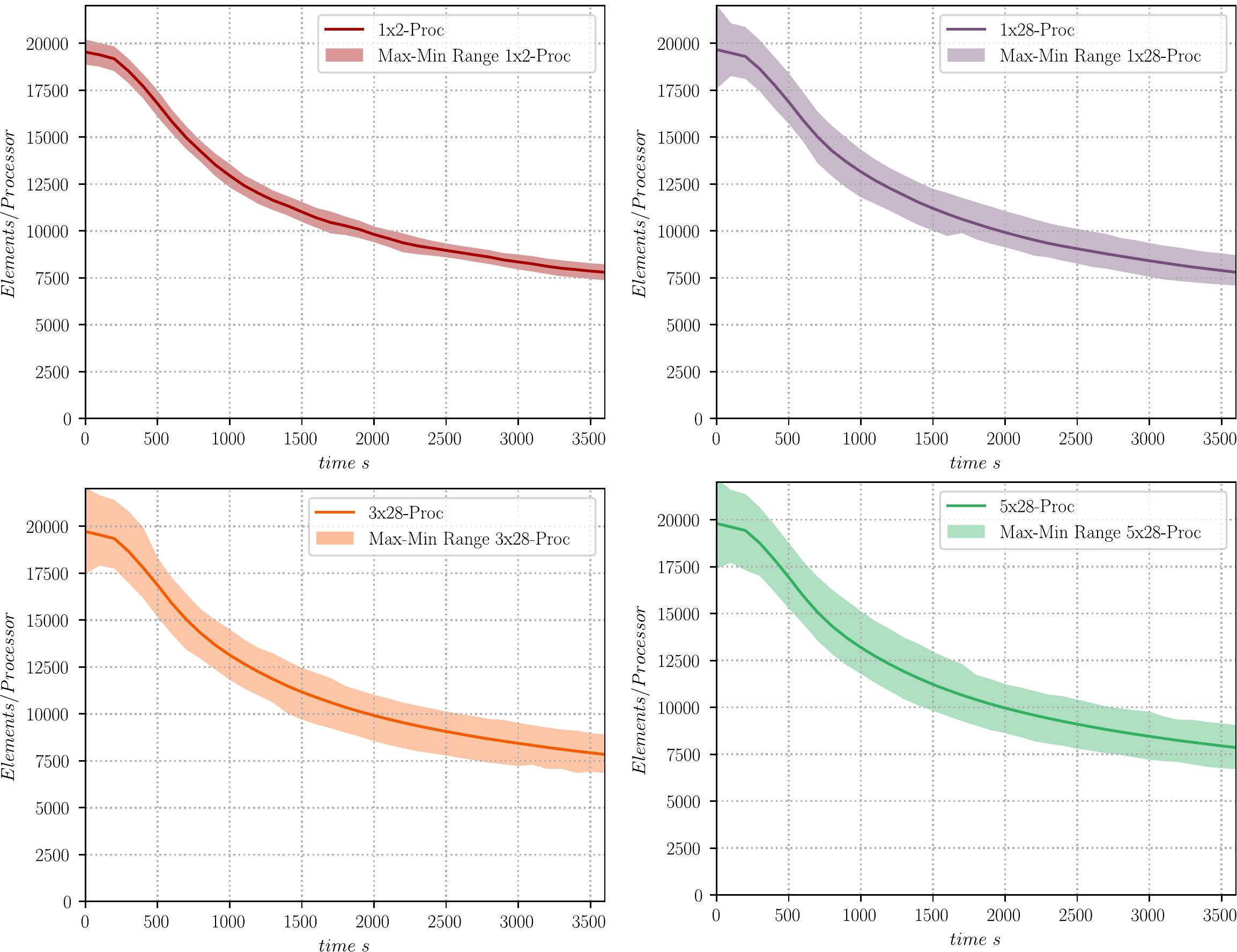}
\caption{Mean number of elements per processor for the test with a DSPP of 2 $mm^2$ performed in 2, 28, 84 (3x28) and 140 (5x28) processors. The range for the number of elements of all processors in the same simulation is shown in the same color with an alpha component.}
\label{fig:MeanElementsAndRange_Progressive}
\end{figure}

\begin{figure}[!h]
\centering
\includegraphics[width=1.0\textwidth] {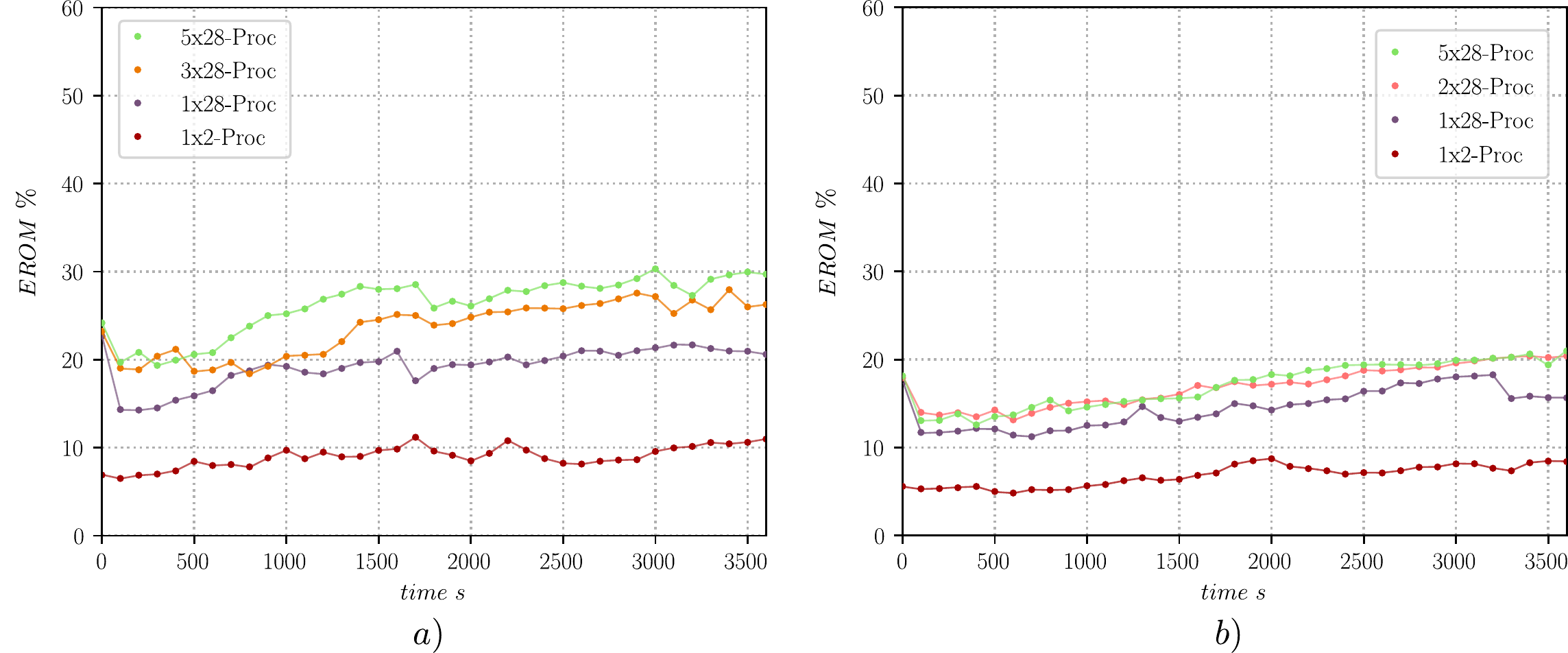}
\caption{Range of the number of elements over the Mean number of elements per processor (index $EROM$) for the test with a DSPP of: a) 2 $mm^2$ and b) 4 $mm^2$ per processor.}
\label{fig:RangePercentPlot_Progressive}
\end{figure}

Figure \ref{fig:MeanElementsAndRange_Progressive} illustrates the evolution of the number of elements for the simulation with a DSPP of 2 $mm^2$, the mean number of elements is very well maintained for all processors although their range in the Y axis increases for the simulations with high number of processors, similarly to the results obtained for the test with a variable processors charge. The evolution of the $EROM$ index is presented in figure \ref{fig:MeanElementsAndRange_Progressive} for both sets of simulations, where the sets using a DSPP of 4 $mm^2$ have a lower overall value, this suggests that the efficiency of these simulations should be higher than the efficiency of those with a DSPP of 2 $mm^2$.\\

\begin{figure}[!h]
\centering
\includegraphics[width=1.0\textwidth] {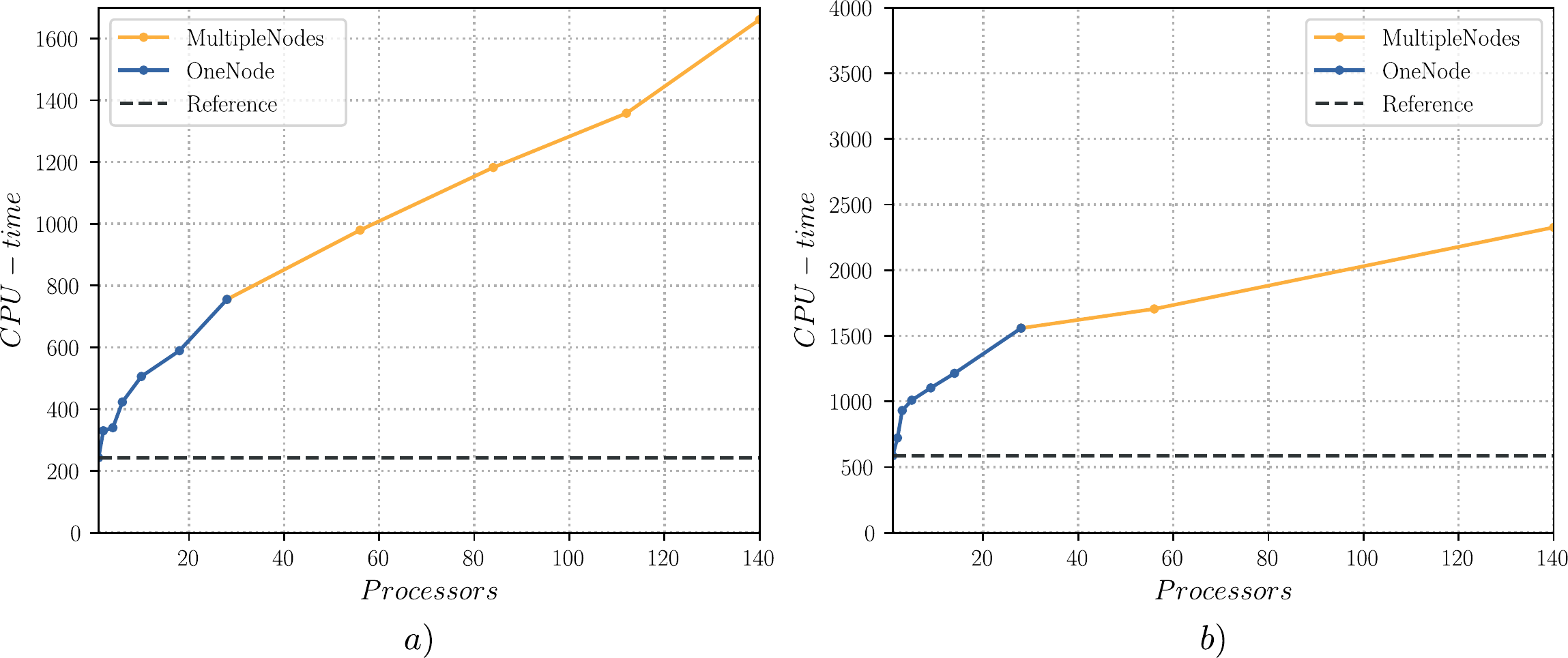}
\caption{CPU-times at the end of the simulation for the test with a DSPP of: a) 2 $mm^2$ and b) 4 $mm^2$ per processor.}
\label{fig:CPU-Time_Progressive}
\end{figure}

\begin{figure}[!h]
\centering
\includegraphics[width=0.8\textwidth] {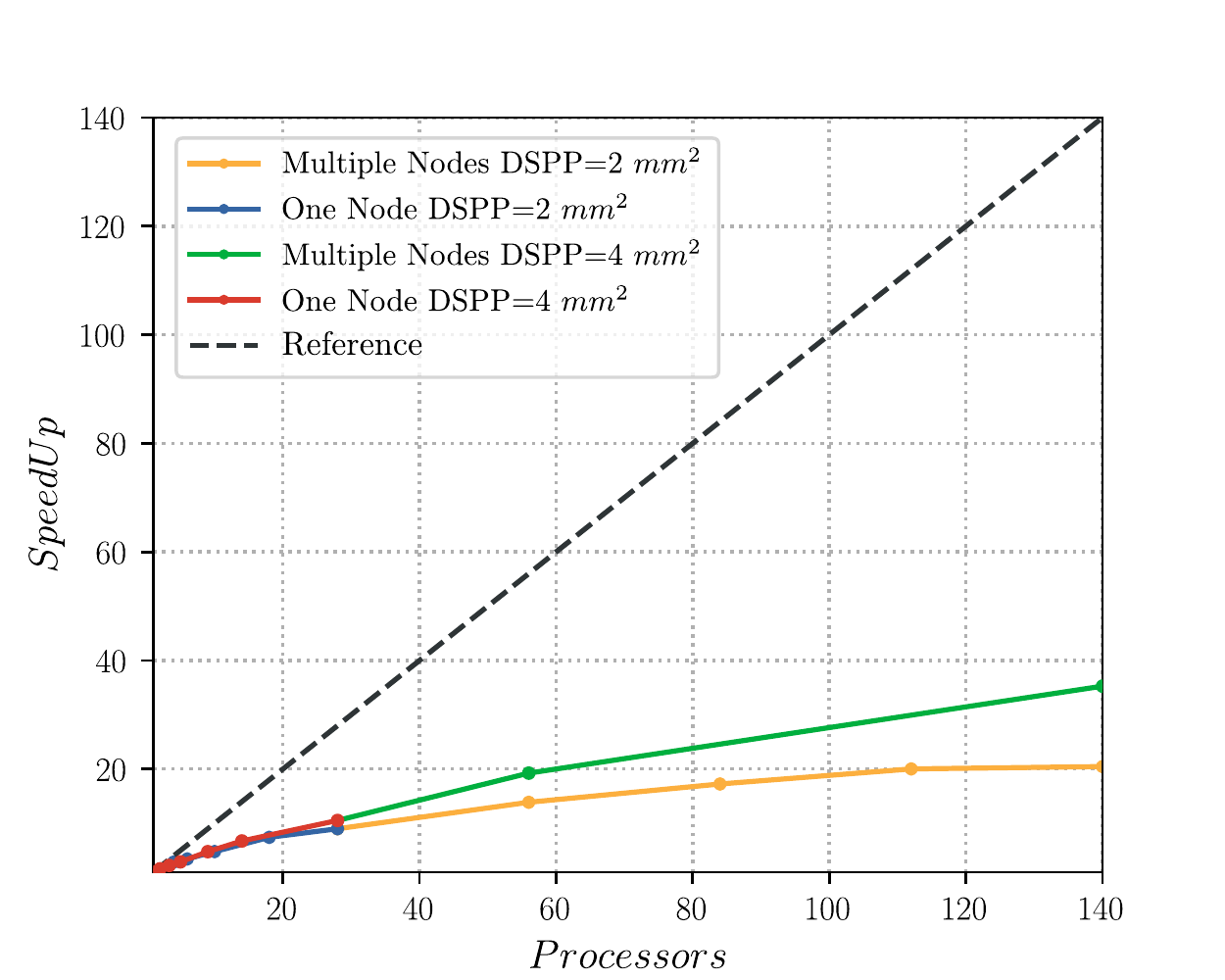}
\caption{Performance of the TRM model in parallel for a constant processor charge: the domain surface increases proportionally to the number of processors, two test were performed, one with a DSPP of 2 $mm^2$ (approximately 2000 grains per processor) and the second with a DSPP of 4 $mm^2$ (approximately 4000 grains per processor). The data is compared to the optimal speed up (Reference).}
\label{fig:SpeedUp_CpuTime_Constant}
\end{figure}

Figures \ref{fig:CPU-Time_Progressive} and \ref{fig:SpeedUp_CpuTime_Constant} show the CPU-time and the speed-up of both sets of simulations. Note than in figure \ref{fig:CPU-Time_Progressive}, the scales have been adapted so the reference curve (the value of the CPU-time for the sequential simulation) be at the same height for both the simulations with a DSPP of 2 $mm^2$ (Fig. \ref{fig:CPU-Time_Progressive}.a)) and of 4 $mm^2$ (Fig. \ref{fig:CPU-Time_Progressive}.b)), hence allowing a better comparison of the relative CPU-time needed for all simulations when compared to their respective reference. clearly the simulations with a DSPP of 4 $mm^2$ need a relatively lower CPU-time to achieve the end of the simulations. This can also be seen in figure \ref{fig:SpeedUp_CpuTime_Constant}: for the simulations run in one node (one single cpu with a maximun 28 processors) the speed-up is very similar, however above 28 processors (for multiple nodes) the speed-up favors the simulations with a higher DSPP.\\

\section{Discussion, conclusion and perspectives}\label{sec:conclusions}

The parallel implementation of the TRM method employed several sub-algorithms each one of them addressing a different problem: the partitioning of the domain, using the dual graph of the initial mesh with the open-source library Metis, and the re-numbering scheme developed to maintain the coherence between partitions generated an initial partitioning. The Mesh scattering algorithm allows re-equilibration of the charges between partitions and performs a 2-step remeshing by blocking the boundaries between partitions and the Lagrangian movement in parallel. Predictions were validated in parallel compared to sequential simulations.\\

Two sets of test cases were studied. The first characterized the performance of the TRM model at ``variable processor charge" (strong scaling benchmark). The accuracy of our model when performed in sequential and parallel was studied, obtaining the same response with negligible errors in every test. The speed-up for these simulations was shown to be dependent on the mean number of elements and on the ``Element Range over Mean" index $EROM$. It was observed that for simulations performed with a high number of processors the speed-up may not be satisfactory if the number of elements is too low or if the $EROM$ index is too high. For a simulation with 600000 initial elements (52983 initial grains) performed over 84 processors, the speed up was lower than for the same simulation performed over 56 processors. These results suggest that this simulation performed with 84 processors is indeed over partitionned and that one should aim to obtain a mean number of elements not lower than 10000 or a $EROM$ index higher than 30\%. Then a test with 544913 initial grains was performed obtaining good results in terms of speed-up for all simulations. The maximum speed-up obtained was for the case with a 544913 initial grains performed over 140 (5x28) processors for which a speed-up of 48.5 was obtained, this simulation took 38 minutes and 45 seconds to perform 360 increments while the one performed in 1 processor took 31 hours and 20 minutes. The total number of grains at the end of this simulation was of 117157 grains, hence 21.5 $\%$ of the total initial number of grains while for the simulation with 52983 initial grains was of 10399, hence 19.6 $\%$ of the total initial number of grains.\\
 
Tests at ``constant processor charge" (weak scaling benchmark) were performed with a simulated domain increasing its size proportional to the number of processors considered. Here a wide range of simulations were performed divided in two sets: the first with a domain surface per processor (DSPP) of 2 $mm^2$ (approximately 2000 grains per processor) and the second with a DSPP of 4 $mm^2$ (approximately 4000 grains per processor). Results of these simulations showed that the speed up obtained for the simulations with a DSPP of 4 $mm^2$ was higher than for the simulations with a DSPP of 2 $mm^2$ as a relatively higher $EROM$ index was found for the latter. \\

Another observation of the tests at ``constant processor charge" was  that the L2-Error over the grain size distributions is obtained to be lower than 3\% after a domain size of 56 $mm^2$, suggesting that our microstructure can be statistically well represented by a simulation with over 56000 initial grains. On the contrary, simulations performed with a domain of 20 $mm^2$ and below, show an error at the end of the simulation of more than 5\% which suggest that they contain too few grains (at the end of the simulation, 3954 grains for the case of with 20000 initial grains) or that statistical results may be highly influenced by the boundary conditions (in our case, orthogonality of grain boundaries with the domain limits), simulations with lower than 3954 grains at the end should be avoided in our GG context. This also enforces the argument behind the studies aiming to increase the performance of massive multidomain simulations as the main obstacle to increase the number of domains in such simulations are given by their high computational cost.\\



In the context of massive multidomain simulations, to the knowledge of the authors, only two methodology in the literature have attempted to perform simulations with hundreds of thousands of domains: the one in \cite{Miesen2017} with a maximun number of 100000 grains and the one in \cite{Elsey2009, Elsey2013} with 671000. Results in \cite{Miesen2017} showed a better speed-up of their parallel implementation than the one presented in this article for our TRM model. Concerning the CPU-time, the parallel strategy presented in \cite{Miesen2017} was able to perform 20 Increments of a simulation with 100000 initial grains in 32 $s$ when performed in 128 processors. In \cite{Elsey2013} no data concerning the speed-up was provided, however, it was mentioned that the simulation with 671000 initial grains was performed until less than 4000 grains remained, with a total CPU-time between 9 and 12 days, when running on 18 intel Nahalem processors. Moreover, main originality of the proposed model here, is, for the first time, to exhibit very efficient simulations in context of unstructured finite element meshes (regular grids in FFT context are considered in \cite{Miesen2017, Elsey2013}). Indeed, a such strategy will enable to consider large deformation of the calculation domain, paving the way to more complex mechanisms such as dynamic recrystallization. Our meshing/remeshing strategy, conserving a description of the bulk of the grains, will also make it possible to investigate mechanisms involved in the grain boundary network, but also, in the grains substructures.\\


The Implementation of this parallel scheme corresponds to the first perspective fulfilled for the general TRM approach. Other perspectives concern: the development of the  a Dynamic Recrystallyzation (DRX)  and Post-Dynamic Recrystallyzation (PDRX) TRM model, the possibility to perform simulation taking into account anisotropic grain boundary properties and the extension of the algorithms to a 3D context. These questions will be discussed in future publications.\\

\section*{Acknowledgements}
The authors thank the ArcelorMittal, ASCOMETAL, AUBERT \& DUVAL, CEA,  FRAMATOME, SAFRAN, TIMET, Constellium and the ANR for their financial support through the DIGIMU consortium and ANR industrial Chair (Grant No. ANR-16-CHIN-0001).

\section*{Data availability}
The raw data required to reproduce these findings cannot be shared at this time as the data also forms part of an ongoing study. The processed data required to reproduce these findings cannot be shared at this time as the data also forms part of an ongoing study.

\bibliography{ms}

\end{document}